\documentclass[pdftex,twocolumn,epjc3]{svjour3}          

\RequirePackage[T1]{fontenc}

\smartqed  

\RequirePackage{graphicx}
\RequirePackage{mathptmx}      
\RequirePackage{flushend}
\RequirePackage[numbers,sort&compress]{natbib}
\RequirePackage[colorlinks,citecolor=blue,urlcolor=blue,linkcolor=blue]{hyperref}
\usepackage{graphicx}
\usepackage{enumerate}
\usepackage{amsmath}
\usepackage{amssymb}
\usepackage{amsfonts}
\usepackage{color}
\usepackage{pstricks}
\usepackage{color}
\usepackage{multirow}
\usepackage[utf8x]{inputenc}
\usepackage{subfigure}

\journalname{Eur. Phys. J. C}

\begin{document}
%
%
\title{Dark matter in Inert Doublet Model with one scalar singlet and $U(1)_X$ gauge symmetry}
%
%
%

\author{M. A. Arroyo-Ure\~na\thanksref{e1,addr1}
        \and
        R. Gaitan\thanksref{e2,addr1} 
        \and
        R. Martinez\thanksref{e3,addr2}
        \and
        J. H. Montes de Oca Yemha\thanksref{e4,addr1}
}

\thankstext{e1}{marcofis@yahoo.com.mx}
\thankstext{e2}{rgaitan@unam.mx}
\thankstext{e3}{remartinezm@unal.edu.co}
\thankstext{e4}{josehalim@comunidad.unam.mx}

\institute{Departamento  de  F\'isica,  FES-Cuautitl\'an, Universidad  Nacional  Aut\'onoma  de  M\'exico,  Estado  de  M\'exico,  M\'exico\label{addr1}
          \and
          Departamento de F\'isica, Universidad Nacional de Colombia, Bogot\'a D.C., Colombia\label{addr2}
}

\date{Received: date / Accepted: date}

\maketitle

%
%
\begin{abstract}
We study Dark Matter (DM) abundance in the framework of the extension of the Standard Model (SM) with an additional $U(1)_X$ gauge symmetry. One complex singlet is included to break the $U(1)_X$ gauge symmetry, meanwhile one of the doublets is considered inert to introduce a DM candidate. The stability of the DM candidate is analyzed with a continuous $U(1)_X$ gauge symmetry as well as discrete $Z_2$ symmetry. We find allowed regions for the free model parameters which are in agreement with the most up-to-date experimental results reported by CMS and ATLAS collaborations, the upper limit on WIMP-nucleon cross section imposed by XENON1T collaboration and the upper limit on the production cross-section of a $Z^{\prime}$ gauge boson times the branching ratio of the $Z^{\prime}$ boson decaying into $\ell^-\ell^+$. We also obtain allowed regions for the DM candidate mass from the relic density reported by the PLANCK collaboration including light, intermediate and heavy masses; depending mainly on two parameters of the scalar potential, $\lambda_{2x}$ and $\lambda_{345}=\lambda_3+\lambda_4+2\lambda_5$. 
We find that trough $pp\rightarrow \chi\chi \gamma$ production, it may only be possible for a future hadron-hadron Circular Collider (FCC-hh) to be able to detect a DM candidate within the range of masses 10-60 GeV.
\end{abstract}
%
%
%
%
%
\section{Introduction}
\label{sec1}
Cosmological observations have shown anomalies that establish the existence of non-luminous matter as a possible solution. This non-luminous matter was called Dark Matter (DM) by F. Zwicky \cite{zwicky}. F. Zwicky applied the virial theorem to the $\textit{Coma Cluster}$ and concluded that a large amount of non-luminous matter must be considered to keep the system bound together. Forty years later, V. Rubin and N. Thonnard found  gravitational evidence through the rotation curve of spiral galaxies \cite{rubin,rubin1,rubin2,rubin3}.
Several proposals arose to explain the observations, namely, modified gravity \cite{milgrom}, a dark component of matter \cite{bradac,bradac1,bradac2,bradac3}, non-baryonic DM \cite{bergstrom}.
Nowadays observations suggest 
the existence of non-baryonic DM as the most viable solution. The PLANCK collaboration reveals that cold non-baryonic content of the matter density is $\Omega h^2=0.120\pm0.001$ \cite{PLANCK}, constituting about $25\%$ of the energy content of the universe.

 As it is well known, the Standard Model (SM) of particle physics \cite{SM,SM1,SM2,Cruz:2019vuo} does not provide answer to fundamental issues; in particular, we can highlight the absence of a DM candidate, which motivates to extend the SM, opening the door to possible new physics beyond SM (BSM).
BSM can include one or more scalar fields to introduce DM candidate, which corresponds to the simplest type of DM known as Weak Interacting Massive Particle (WIMP) \cite{IDM1, IDM2, IDM3, IDM4, IDM5, SUSYDM, Bertone1, Bertone2}.
The scalar particle as DM candidate must satisfy experimental and theoretical constraints \cite{10testpoint}; for instance: it must have the right relic density, neutral particle and it must be consistent with direct DM searches. On the experimental side, searches for WIMPs based on different methodologies are realized by collaborations such as given in Refs. \cite{DirectDM,DirectDM1,DirectDM2}, the CDMS \cite{cdms}, CoGeNT \cite{cogent}, Xenon \cite{xenon} and LUX \cite{lux}. The second one is through indirect searches \cite{IndirectDM} by PAMELA \cite{pamela}, ATIC \cite{atic}, and Fermi LAT \cite{fermilat} experiments for particles resulting from WIMP annihilation, for example, positron-electron pairs.
Finally, DM search at colliders \cite{DMcollidersATLAS}, \cite{DMcollidersCMS}, such as the LHC, the WIMPs can be produced in pairs in association with other particles. A process to study DM at colliders is $pp\to \chi\chi+P$, where $\chi$ is a DM candidate and $P=g,\,\gamma,\,W,\,Z,\,H$. 

The simplest proposal for a DM candidate is to extend the SM by introducing a singlet scalar field \cite{McDonald:1993ex}. An interesting and simple model with a scalar field as DM candidate is the Inert Higgs Doublet Model (IDM) \cite{Deshpande:1977rw} which contains a neutral scalar particle to play the role of WIMP \cite{Dolle:2009fn}. The IDM shows an important dependence on the mass splitting parameter defined as the masses difference between pseudo-scalar and scalar coming from second doublet, the inert doublet. The heavy DM mass region for small values of the mass splitting parameter is obtained for masses from 500 GeV to 1000 GeV, meanwhile, the light DM mass region for the mass splitting parameter of the order of 50 GeV - 90 GeV is obtained for masses from 30 GeV to 80 GeV. 

Other possibilities are Supersymmetry, which provides a WIMP candidate through the lightest neutralino \cite{susy,susy1}, or universal extra dimension models with the lightest Kaluza-Klein partner as DM candidate \cite{ED,ED1,ED2,ED3,ED4}. 

In this work, we consider a model with an additional $U(1)_X$ gauge symmetry which includes two doublets and one complex singlet of scalar fields. One doublet is inert of which we identify a degree of freedom as
a DM candidate, meanwhile, the other doublet is the usual SM doublet.
The stability of the DM candidate is ensured by imposing a discrete $Z_2$ symmetry or by $U(1)_X$ gauge symmetry. Models with extra $U(1)_X$ gauge symmetries as extensions of the SM has many motivations. For example, Grand Unified and superstring theories contain additional $U(1)_X$ factors in the effective low energy limit. Supersymmetric extensions include theoretical and phenomenological aspects such as flavor physics, neutrino physics and DM \cite{Hewett:1988xc, Suematsu:1994qm, Demir:2005ti, Langacker:2008yv}. Extended models with a $U(1)_X$ gauge symmetry also have phenomenological importance because they predict a heavy vector gauge boson $Z^{\prime}$ derived from the spontaneous symmetry breaking (SSB) \cite{Martinez:2014rea}, \cite{Camargo:2018klg}. Besides, the $U(1)_X$ gauge symmetries can be incorporated in  extended models that are free from triangle anomalies adding new fermions.

Previously, one of our authors published a paper with a similar approach, it can be found in the Ref. \cite{Martinez:2014ova}, in which DM candidate mass of the order of $1.3$ GeV to $70$ GeV are allowed, depending on the assignment of the free parameters associated. The experimental data from LEP and relic density observation are considered to find an allowed mass of DM candidate of the order of 70 GeV in a scenario that assigns the parameters of the model as Higgs-phobic type, in which the $Z^\prime$ boson provides the channel of annihilation for DM suppressing the participation of the Higgs channel. The decay signal of Higgs diphoton also imposes strong restrictions through recent data from the CERN-LHC collider \cite{Martinez:2015kmn}. When it is combined with the observed value of DM relic density, the allowed mass region is obtained such that $5$ GeV $\leq m_{\chi}\leq 62$ GeV for values of the order of $0.02\sim 0.08 $ of the quartic coupling between doublets and singlet scalar in a model with $U(1)$ gauge symmetry \cite{Martinez:2014rea}. In reference \cite{DMseveral,DMseveral1,DMseveral2,DMseveral3,DMseveral4,DMseveral6,DMseveral7,DMseveral8,DMseveral9,DMseveral10} we include extensive literature to be consulted.

The organization of our research is as follows. In Sec. \ref{sec2} we give a general view of the model.  Sec. \ref {sec3} is focused to constrain the free model parameters. In Sec. \ref{sec4}, we show the branching ratios for the $Z^{\prime}$ and the neutral scalar associated with the singlet field. Sec. \ref{sec5} is devoted to the analysis of the relic density, we present our results and an analysis of them.  The DM production at future colliders through $pp\to\chi\chi\gamma$ is presented in Sec. \ref{sec6}. Finally, in Sec. \ref{sec7} conclusions are presented.
\section{Inert Doublet Model plus a complex Singlet scalar (IDMS)}
\label{sec2}
The IDMS incorporates a local $U(1)_X$ gauge symmetry and a $SU(2)$ scalar doublet to the SM gauge symmetry 
$G_{\text{SM}}=SU(3)_C\otimes SU(2)_L \otimes U(1)_Y$. 
The $Z^\prime$ gauge boson associated with $U(1)_X$ will provide an additional channel to the production and annihilation in scattering processes. 
On the other hand, the singlet scalar field is included to break down the $U(1)_X$ symmetry to $G_{\text{SM}}$. The DM candidate arises from the second doublet scalar field, which has a Vacuum Expectation Value (VEV) equal to zero to guarantee the stability of the DM candidate. But not only with a null value of VEV can achieve the stability of DM candidate, but it is also necessary a mechanism to control the couplings responsible for the DM candidate decays. Two possible options to control the stability of the DM are considered: a discrete $Z_2$ symmetry or the $U(1)_X$ gauge symmetry \cite{Haber:2018iwr,Bonilla:2014xba,Krawczyk:2015xhl}. 
\subsection{Scalar fields}
The scalar fields and their assignments under the $G_{\text{SM}}\otimes U(1)_X $ group are given by:
\begin{eqnarray}
\Phi_1&\sim&(\textbf{1}, \textbf{2}, 1/2, x_1),\\ \nonumber
\Phi_2&\sim&(\textbf{1},\textbf{2},1/2,x_2),\\ \nonumber
\mathcal{S}_X&\sim&(\textbf{1},\textbf{1},0,x),
\end{eqnarray}
where two first entries denote the representation under $SU(3)_C$ and $SU(2)_L$, respectively, meanwhile the hypercharge and charge under $U(1)_X$ are written in the last two entries.
The scalar fields are written as follows:
\begin{eqnarray}
\Phi_1 &=& \left( \begin{array}{c}
\phi_1^+\\  \nonumber
\frac{1}{\sqrt{2}}(\upsilon+\phi_1+i\eta_1)
\end{array} \right),\\
\Phi_2 &=& \left( \begin{array}{c}
\phi_2^+\\
\frac{1}{\sqrt{2}}(\phi_2+i\eta_2)
\end{array} \right),\\ \nonumber
\mathcal{S}_X &=& \frac{1}{\sqrt{2}}(\upsilon_x+s_x+i\eta_x).
\label{eq:scalar_reps}
\end{eqnarray}

The spontaneous symmetry breaking (SSB) is achieved as
\begin{equation*}
G_{\text{SM}}\otimes U(1)_X  \xrightarrow{\langle \mathcal{S}_X \rangle} G_{\text{SM}} \xrightarrow{\langle \Phi_1 \rangle} SU(3)_C\otimes U(1)_{\text{EM}},
\end{equation*}
where $\langle \mathcal{S}_X \rangle = \upsilon_x/\sqrt{2}$ and  $\langle \Phi_1 \rangle^T = (0,\upsilon/\sqrt{2})$ with $\upsilon=246$~GeV. Note that $\Phi_2$ must have VEV equal to zero to guarantee the stability of the DM candidate.
The most general, renormalizable and gauge invariant potential is
\begin{eqnarray}
V &=&\mu _{1}^{2}\Phi _{1}^{\dag }\Phi _{1}+\mu _{2}^{2}\Phi _{2}^{\dag}\Phi _{2}+\mu _{x}^{2}\mathcal{S}_{X}^{\ast }\mathcal{S}_{X}+\left[ \mu _{12}^{2}\Phi_{1}^{\dag }\Phi _{2}+h.c.\right]  \nonumber \\
&+&\lambda _{x}\left( \mathcal{S}_{X}^{\ast }\mathcal{S}_{X}\right) ^{2}+\lambda _{1}\left( \Phi _{1}^{\dag }\Phi _{1}\right) ^{2}+\lambda_{2}\left( \Phi _{2}^{\dag }\Phi _{2}\right) ^{2} \nonumber \\
&+&\lambda _{3}\left( \Phi_{1}^{\dag }\Phi _{1}\right) \left( \Phi _{2}^{\dag }\Phi _{2}\right)+\lambda _{4}\left\vert \Phi _{1}^{\dag }\Phi _{2}\right\vert ^{2}+ \left[\lambda _{5}\left( \Phi _{1}^{\dag }\Phi _{2}\right) ^{2}\right.\nonumber \\
&+&\left. \lambda _{6}\left(\Phi _{1}^{\dag }\Phi _{1}\right) \left( \Phi _{1}^{\dag }\Phi _{2}\right)
\right.   \left. +\lambda _{7}\left( \Phi _{2}^{\dag }\Phi _{2}\right) \left( \Phi_{1}^{\dag }\Phi _{2}\right) +h.c.\right]   \nonumber \\
&+&\left( \mathcal{S}_{X}^{\ast }\mathcal{S}_{X}\right) \left[ \lambda _{1x}\left( \Phi_{1}^{\dag }\Phi _{1}\right) +\lambda _{2x}\left( \Phi _{2}^{\dag }\Phi_{2}\right) \right]  \nonumber \\
&+&\left[ \lambda _{12x}\left( \Phi _{1}^{\dag }\Phi _{2}\right) \left( \mathcal{S}_{X}^{\ast }\mathcal{S}_{X}\right) +h.c.\right],
\label{potential}
\end{eqnarray}
where $\mu_{1,\,2}^2$, $\lambda_{1,\,2,\,3,\,4,\,1x,\,2x}$ are real parameters and  $\mu_{12}^2$, $\lambda_{5,\,6,\,7,\,12x}$ can be complex parameters. Note that $\mu_2^2>0$ because the DM candidate arises from $\Phi_2$, which has $<\Phi_2> = 0$. The terms in the scalar potential that are proportional to $\Phi _{1}^{\dag }\Phi _{2} \mathcal{S}_{X}$ or $\Phi _{2}^{\dag }\Phi _{1} \mathcal{S}_{X}$ can generate a decay of DM candidate into two neutral scalars. We assume that $x_2-x_1\pm x\neq0$ to leave these terms non-invariant under gauge symmetry. Thus, the parameters that accompany these terms must be zero to recover the gauge invariance and at the same time eliminate the couplings that are responsible for a decay of DM candidate at two neutral scalars.

After SSB the mass matrix for scalars in the $ \{ \phi_1, \,s_x,\, \phi_2,\, \eta_2 \}$ basis is
\begin{equation}
M_{0}^2=\left(\begin{array}{cccc}M_{11} & M_{12} &M_{13} &0 \\M_{12} & M_{22} & M_{23} &0\\M_{13} & M_{23} & M_{33} & M_{34} \\0 & 0 & M_{34} & M_{44}\end{array}\right),
\label{mass0}
\end{equation}
where
\begin{eqnarray}
M_{11}&=&2\lambda_1\upsilon^2,\;\;\;\;\;   M_{12}=\lambda_{1x}\upsilon \upsilon_x,  \;\;\;\;\; M_{13}=\frac{1}{2}\lambda_{6}\upsilon^2,\nonumber\\
M_{22}&=&2\lambda_{x}\upsilon_x^2, \;\;\;\;\;M_{23}=\frac{1}{2}\lambda_{12x}\upsilon \upsilon_x,\nonumber\\ 
M_{33}&=&\mu_2^2+\frac{1}{2}(\lambda_3+\lambda_{4}+\textrm{Re}[\lambda_{5}])\upsilon^2+\frac{1}{2}\lambda_{2x}\upsilon_x^2,\nonumber\\
M_{34}&=&-\textrm{Im}[\lambda_{5}]\upsilon^2,\nonumber \\
M_{44}&=&\mu_2^2+\frac{1}{2}(\lambda_3+\lambda_{4}-\textrm{Re}[\lambda_{5}])\upsilon^2+\frac{1}{2}\lambda_{2x}\upsilon_x^2.
\label{massmatrix}
\end{eqnarray}
After the $M_{0}^2$ matrix is diagonalized and neutral scalars are rotated to physical states, the $M_{13}$ and $M_{23}$ matrix elements allow the mixing between neutral scalars and DM candidate, as shown in equation (4). This means that terms proportional to $ \Phi _{1}^{\dag }\Phi _{2}$ in the scalar potential must be eliminated, otherwise DM candidate will be unstable. 
For $x_1=x_2$ the terms proportional to $ \Phi _{1}^{\dag }\Phi _{2}$ in the scalar potential are gauge invariant. Then, it is required to introduce an additional discrete $Z_2$ symmetry for the doublets to eliminate these terms in the potential. Moreover, for $x_1\neq x_2$ the gauge invariance of the $U(1)_X$ symmetry guarantees the stability for the DM candidate. In either case, we will assume that $\lambda_6=\lambda_7=\lambda_{12x}=0$ in order to maintain the invariance under $Z_2$ or $U(1)_X$ symmetries. 
\subsection{$Z_2$ symmetry and $x_1=x_2$ case}
The terms proportional to $\Phi_{1}^{\dag }\Phi _{2}$ in the potential are invariant under $U(1)_X$; then it is necessary to introduce a $Z_2$ discrete symmetry to eliminate them. The proper assignment is $\Phi_1\rightarrow \Phi_1$ and $\Phi_2\rightarrow -\Phi_2$. Under the last assignment for the doublet, the $M_{13}=M_{23}=0$ and the mass matrix for the neutral scalar, Eq. (\ref{mass0}), can be diagonalized by
\begin{equation}
\left(\begin{array}{c} h \\ S\end{array}\right)=\left(\begin{array}{cc}\cos\alpha_1 & -\sin\alpha_1 \\ \sin\alpha_1 & \cos\alpha_1
\end{array}\right) \left(\begin{array}{c} \phi_1 \\ s_x \end{array}\right)
\label{alpha1}
\end{equation}
and
\begin{equation}
\left(\begin{array}{c} \chi \\ A\end{array}\right)=\left(\begin{array}{cc}\cos\alpha_2 & -\sin\alpha_2 \\ \sin\alpha_2 & \cos\alpha_2
\end{array}\right) \left(\begin{array}{c} \phi_2 \\ \eta_2 \end{array}\right),
\end{equation}
where $\tan\alpha_{1,2}=\frac{r_{1,2}}{1+\sqrt{1+r_{1,2}^2}}$ with $r_1=\frac{\lambda_{1x}\upsilon \upsilon_x}{\lambda_1\upsilon^2-\lambda_{x}\upsilon_x^2}$ and $r_2=\frac{-\textrm{Im}[\lambda_5]}{\textrm{Re}[\lambda_5]}$ \cite{Cabral-Rosetti:2017mai}. Therefore the masses for the scalars are
\begin{equation}
\label{HiggsBosonMass}
m_{S,h}^2=\lambda_1\upsilon^2+\lambda_{x}\upsilon_x^2\pm(\lambda_1\upsilon^2+\lambda_{x}\upsilon_x^2)\sqrt{1+r_1^2},
\end{equation}
while the $H^\pm$ charged scalar, $A$ pseudoscalar and  $\chi$ masses are given, respectively, by
\begin{eqnarray}
m_{H^\pm}^2&=&\mu_2^2+\frac{1}{2}(\lambda_3 \upsilon^2+\lambda_{2x}\upsilon_x^2)\label{eqmCHmass},\\
m_{A,\chi}^2&=&m_{H^{\pm}}^2+\left(\frac{\lambda_4}{2}\pm|\lambda_5|\right) \upsilon^2\label{mass34}.
\end{eqnarray}
We assume that $\chi$ plays the role of DM. 

Two interesting limits can arise when approximations are realized about the $\lambda_5$ quartic couplings involved in $r_{2}$. The LHC results imply that $\upsilon\ll \upsilon_x$ and $\lambda_{1x}\sim 1$, then  $r_1\approx-\frac{\upsilon}{\upsilon_x}$ and $\tan\alpha_1\approx -\frac{\upsilon}{2\upsilon_x}$. By considering $\textrm{Im}[\lambda_5]\sim\textrm{Re}[\lambda_5]$, then $r_2\approx -1$ and $\tan\alpha_2\approx -\frac{1}{1+\sqrt{2}}$. Moreover, if $\textrm{Im}[\lambda_5]=0$, which is the CP conservation case, then $r_2=0$ and $\tan\alpha_2=0$. By considering the previous approximation on $r_1$ and $\tan\alpha_1$, we can write Eq. (\ref{HiggsBosonMass}) as
\begin{eqnarray}
m_h^2&\approx& 2\lambda_1 \upsilon^2, \nonumber\\ 
m_S^2&\approx& 2\lambda_{x} \upsilon_x^2.
\end{eqnarray}

An important fact is that the model allows the $\chi\to H^{\pm}W^{\mp}$ decay, whose $\chi H^{\pm}W^{\mp}$ coupling is shown in table \ref{FR-IDMS}. To avoid the instability of the DM candidate, we demand that the masses must satisfy $m_{H^\pm}^2\,(m_{A}^2)>m_{\chi}^2$. To achieve this, from eqs. (\ref{eqmCHmass}) and (\ref{mass34}), we impose the following constraint: $\lambda_4 > 2|\lambda_5|$.

\subsection{$U(1)_X$ gauge symmetry, $x_1\neq x_2$ case}
The DM candidate can also be stable when $x_1\neq x_2$. In this case, the same parameters in the scalar potential, as the previous case, must be eliminated and $\lambda_5$ must be also zero.
In addition, $\phi_2$ and $\eta_2$ are not mixing since $M_{34}=0$. 
\subsection{Gauge bosons interactions}
The kinetic terms for the $U(1)_Y$ and $U(1)_X$ gauge symmetries are given by:
\begin{equation}
\mathcal{L}_{Kin} = - \frac{1}{4} \hat{B}_{\mu\nu} \hat{B}^{\mu\nu}+ \frac{1}{2}\frac{\varepsilon}{\cos\theta_{W}} \hat{B}^{\mu\nu} \hat{Z}_{0\mu\nu}^{\prime}
- \frac{1}{4}\hat{Z}_{0\mu\nu}^{\prime}\hat{Z}^{\prime 0\mu\nu},
\label{eq:kin}
\end{equation}
where, $ \hat{B}^{\mu\nu}$ and  $\hat{Z'_0}^{\mu\nu}$ are the field strength tensors defined by $\hat{F}_{\mu\nu}=\partial_\mu \hat{F}_\nu-\partial _\nu \hat{F}_\mu$ for $\hat{F}_{\nu}=\hat{B}_{\nu},\,\hat{Z}_{0\nu}$ \cite{Lee:2013fda,Davoudiasl:2012ag}. The mixing term between $\hat{B}_{\mu\nu}$ and $\hat{Z}_{0\mu\nu}^{\prime}$ is allowed by the gauge invariance. However, this mixing term can be eliminated by the field redefinition
\begin{eqnarray}
 \left( \begin{array}{c}
 Z'_{0\mu}\\
B_\mu
\end{array} \right)
=
 \left( \begin{array}{cc}
\sqrt{1-\varepsilon^2/\cos^2\theta_W} & 0\\
-\varepsilon/ \cos^2\theta_W&1 \end{array} \right) \left( \begin{array}{c}
 \hat{Z'}_{0\mu}\\
\hat{B}_\mu
\end{array} \right),
\end{eqnarray}
where the fields with hat notation contain the kinetic mixing term and $\varepsilon$ must be small to agree with the experiment. After SSB, the gauge bosons in the mass basis are
\begin{equation}
A_\mu=\hat{A}_\mu-\varepsilon \hat{Z}_{0\mu}^{\prime},
\end{equation}
\begin{equation}
Z_{0\mu}=\hat{Z}_{0\mu}+\varepsilon\tan\theta_W \hat{Z}_{0\mu}^{\prime},
\end{equation}
\begin{equation}
Z_{0\mu}^\prime=\hat{Z}_{0\mu}^{\prime}.
\end{equation}
The parameter $\varepsilon$ is assumed to be $\varepsilon\ll\cos\theta_W$ in order to ignore terms higher or equal to $\mathcal{O}(\varepsilon^2)$. The term $\varepsilon$ is constrained experimentally with values smaller than $10^{-3}$ \cite{Abel:2008ai}.

The interaction between gauge and scalar fields is
\begin{eqnarray}
\mathcal{L}_\text{scalar} &=& | D_\mu \Phi_1 |^2 + | D_\mu \Phi_2 |^2 + | D_\mu \mathcal{S}_X |^2,
\end{eqnarray}
where the covariant derivative $D_{\mu}$ for neutral gauge bosons is defined as
\begin{eqnarray}
D_{\mu} =  \left(  \partial_\mu + i g^{\prime} Y \hat B_\mu + i g T_3 \hat W_{3 \mu} + i g_{x} Q^{\prime}_i \hat Z^{\prime}_{0\mu} \right) ,
\end{eqnarray}
where $g_x$ and $Q^{\prime}_i$ are the coupling constants and the charge for $U(1)_X$, respectively. When the SSB is achieved not only the mass terms are generated but also mixing terms are obtained:
\begin{equation}
\mathcal{L}_\text{scalar}=\frac{1}{2}m_{Z^\prime}^2Z^{\prime0} Z^{\prime0}+\frac{1}{2}m_{Z}^2Z^0Z^0-\Delta^2Z^{0} Z^{\prime0}+...,
\end{equation}
where
\begin{eqnarray}
m_{Z^\prime}^2 &=& \left( \frac{g^\prime\varepsilon}{2\cos\theta_W}+g_x x_1 \right)^2 \upsilon^2+g_x^2x^2\upsilon_x^2\\ \nonumber
&\approx& g_x^2 x^2 \upsilon_x^2, 
\end{eqnarray} \label{eq:mZ'}
and
\begin{equation}
\Delta^2=\frac{1}{2}g_Z\left(\frac{g^\prime\varepsilon}{2 \cos\theta_W}+g_x x  \right)\upsilon^2,
\end{equation}
meanwhile, the $Z$ gauge boson mass retains the same value set by the SM,
\begin{equation}
m_{Z}^2 = g^2\frac{\upsilon^2}{4\cos^2\theta_W}.
 \label{eq:mZ}
\end{equation}
In order to cancel the mixing term, the following rotation is required
\begin{eqnarray}
\left(  \begin{array}{c}
Z \\ Z^\prime
\end{array} \right)
=
\left( \begin{array}{cc}
\cos\xi & -\sin\xi \\ \sin\xi & \cos\xi
\end{array} \right)
\left(
\begin{array}{c}  Z^0 \\ Z^{\prime0}
\end{array} \right),
\end{eqnarray}
where the mixing angle $\xi$ satisfies the expression $\tan 2\xi = \frac{2\Delta^2}{m^2_{Z^0}-m^2_{Z^{'0}}}$, and has been constrained to the interval $|\xi|<10^{-3}$ \cite{Bouchiat:2004sp}.
%
%
\subsection{Fermion interactions}
The most general Yukawa Lagrangian is
\begin{eqnarray}
\mathcal{L}_{Yukawa}=&&\sum_{i,j=1}^{3}\sum_{a=1}^{2}\left( \overline{q}
_{Li}^{0}Y_{aij}^{0u}\widetilde{\Phi }_{a}u_{Rj}^{0}+\overline{q}
_{Li}^{0}Y_{aij}^{0d}\Phi _{a}d_{Rj}^{0} \right.\nonumber \\
&&+\left.\overline{l}_{Li}^{0}Y_{aij}^{0l}
\Phi _{a}e_{Rj}^{0}+h.c.\right) ,  \label{yukawa}
\end{eqnarray}
where $Y_{a}^{0f}$ are the $3\times 3$ Yukawa matrices, for $f=u,d,l$. $q_{L}$ and $l_{L}$ denote the left-handed fermion doublets under $SU(2)_L$, while $u_{R}$, $d_{R}$, $l_{R}$ correspond to the right-handed singlets. The zero superscript in fermion fields stands for the interaction basis. The DM stability is lost if the couplings $Y_{2ij}^{0f}$ appear in the Eq. (\ref{yukawa}). These Yukawa couplings can be eliminated by the correct assignment of values for charges under the $Z_2$ and $U(1)_X$ symmetries, as previously done.

In the case of discrete $Z_2$ symmetry with $x_1=x_2$, the couplings $Y_{2ij}^{0f}$ must be equal to zero in order to respect the discrete $Z_2$ symmetry. The couplings $Y_{1ij}^{0f}$ are allowed if the assignment of the $U(1)_X$ charges for the fermions satisfy  
\begin{equation}
\mp x_1-x_q+x_{u,d}=0
\label{xquark}
\end{equation}
and
\begin{equation}
x_1-x_l+x_{e}=0
\label{xlepton}
\end{equation}
where $x_{q,l}$ are the $U(1)_X$ charges of left-handed doublet fermions, meanwhile, $x_{u,d,e}$ are the $U(1)_X$ charges of right-handed fermions.

In the case of $x_1\neq x_2$ we set the $U(1)_X$ charges such that $\mp x_2-x_q+x_{u,d}\neq0$ and  $x_2-x_l+x_{e}\neq0$ in order to eliminate the couplings $Y_{2ij}^{0f}$ in Eq. (\ref{yukawa}). Obviously,  $\Phi_1$ also satisfies Eqs. (\ref{xquark}) and (\ref{xlepton}) to provide the masses of the fermions as in SM. Feynman rules of IDMS are shown in table \ref{FR-IDMS}.

It is important to mention that the fermion charges under $U(1)_X$ must satisfy the triangle anomaly equations, which can be reviewed in \cite{Mantilla:2016lui}, in order to garantize an anomaly free model. The anomaly cancellation requirements for fermion charges $x_{f}$, for $f=q,u,d,l,e$, are shown in table \ref{table_fermion_x} as a function of $x_q$.

\begin{table}
\caption{IDMS couplings involved in the calculations of this work.
We define $\lambda_{345}=\lambda_{3}+\lambda_{4}+2\lambda_{5}$. For $Z^{\prime}f_i\bar{f}_i$ coupling we consider the limit when the kinetic mixing term $\varepsilon\to0$.\label{FR-IDMS}}
\centering{}%
\begin{tabular}{c c}
\hline
Coupling & Expression\tabularnewline
\hline
\hline
$hf_{i}\bar{f}_{i}$ & $\frac{m_{f_{i}}}{\upsilon}\cos\alpha_{1}$\tabularnewline
\hline
$hH^{-}H^{+}$ &$(\lambda_3\cos\alpha_{1}+\lambda_{2x}/2)\upsilon$    \tabularnewline
\hline
$hW_{\mu}^{-}W_{\nu}^{+}$ & $gm_{W}\cos\alpha_{1}g_{\mu\nu}$\tabularnewline
\hline
$h\chi\chi$ & $(\lambda_{345}\cos\alpha_{1}+\lambda_{2x}/2)\upsilon$\tabularnewline
\hline
$Z^{\prime}_{\mu}f_{i}\bar{f}_{i}$ & $\frac{g_x}{2}\left(1-\gamma^{5}\right)\gamma^{\mu}$\tabularnewline
\hline
$Z_{\mu}^{\prime}\chi\chi$ & $\frac{g_x}{2}(p_{Z_{\mu}^{\prime}}-p_{\chi})^{\mu}$\tabularnewline
\hline
$Sf_{i}\bar{f}_{i}$ & $\frac{m_{f_{i}}}{\upsilon}\sin\alpha_{1}$\tabularnewline
\hline
$SW^-_{\mu}W^+_{\nu}$ & $gm_W\sin\alpha_{1}g_{\mu\nu}$\tabularnewline
\hline
$S\chi\chi$ & $\lambda_{2x}\upsilon_{x}\cos\alpha_{1}-\lambda_{345}\upsilon\sin\alpha_1$\tabularnewline
\hline
$\chi (A) H^{\pm}W_{\mu}^{\mp}$ & $i \frac{g}{\sqrt{2}}(p_{H^{\pm}}-p_{\chi(A)})^{\mu}\cos\alpha_2$\tabularnewline
\hline
\end{tabular}
\end{table}
\begin{table}
	\caption{Relations between fermions charges under $U(1)_X$ to guarantee the anomaly cancellation.}
	\label{table_fermion_x}
	\centering{}%
	\begin{tabular}{c c}
		\hline
		Field & $U(1)_X$\tabularnewline
		\hline
		\hline
		$q_L$ &  $x_q$\tabularnewline
		\hline	
		$u_R$ & $x_u=4x_q$ \tabularnewline
		\hline
		$d_R$ &  $x_d=-2x_q$ \tabularnewline
		\hline
		$l_L$ &  $x_l=-3x_q$ \tabularnewline
		\hline
		$e_R$ &  $x_e=-6x_q$ \tabularnewline
		\hline
	\end{tabular}
\end{table}
%
%



\section{Constraints on free model parameters}
\label{sec3}
In this section, we obtain the experimentally allowed regions for the free model parameters involved in our analysis by considering the most up-to-date experimental collider results reported by CMS \cite{Sirunyan:2018koj} and ATLAS \cite{ATLAS:2018doi} collaborations, namely, signal strengths, denoted by $\mathcal{R}_{x\bar{x}}$. In this work we consider the production of $H_i$ via gluon fusion and we use the narrow width approximation. Then, $\mathcal{R}_{x\bar{x}}$ can be written as follows:
\begin{equation}
\mathcal{R}_{x\bar{x}}\approx\frac{\Gamma(h^{\text{IDMS}}\to gg)\cdot\mathcal{B}(h^{\text{IDMS}}\to x\bar{x})}{\Gamma(h^{\text{SM}}\to gg)\cdot\mathcal{B}(h^{\text{SM}}\to x\bar{x})}
\end{equation}
where $\Gamma(H_i\to gg)$ is the decay width of $H_i$ into gluon pair, with $H_i=h^{\text{IDMS}}$ and $h^{\text{SM}}$. Here $h^{\text{IDMS}}$ is the SM-like Higgs boson coming from IDMS and $h^{\text{SM}}$ is the SM Higgs boson; $\mathcal{B}(H_i\to x\bar{x})$ is the branching ratio of $H_i$ decaying into a $x\bar{x}$, where $x\bar{x}=b\bar{b},\;\tau^-\tau^+,\;\mu^-\mu^+,\;WW^*,\;ZZ^*,\;\gamma\gamma$.
Besides to measurements of colliders, we use the most-up-date upper limit on WIMP-nucleon cross-section, for the spin independent case, reported by XENON1T collaboration \cite{Aprile:2018dbl} and whose value for a DM candidate mass of $30$ GeV is given by:
 \begin{equation}
 \sigma^{SI}(\chi N\to \chi N)< 4.1\times 10^{-47} \text{cm}^2=4.1\times 10^{-7} \text{pb}.
\end{equation}

 On the other side, the free parameters of the IDMS involved in our analysis are the following:
\begin{itemize}
\item Mixing angle $\alpha_{1}$.
\item Vacuum Expectation Value of the scalar singlet, $\upsilon_x$.
\item $U(1)_X$ coupling constant, $g_x$.
\item $Z^{\prime}$ gauge boson mass, $m_{Z^{\prime}}$.
\item Scalar mass, $m_S$.
\item Charged scalar boson mass, $m_{H^{\pm}}$.
\item Dark matter boson mass, $m_{\chi}$.
\end{itemize}

In order to constrain the ${Z^{\prime}}$ gauge boson mass, $m_{Z^{\prime}}$, the upper limit on the production cross-section of a $Z^{\prime}$ gauge boson times the branching ratio of the $Z^{\prime}$  decaying into $\ell^-\ell^+$ \cite{Aaboud:2017buh}, with $\ell=e,\,\mu$, was considered. 


\subsection{Constraint on mixing angle $\alpha_1$}
\begin{figure}[!h]
\center{\includegraphics[scale=0.125]{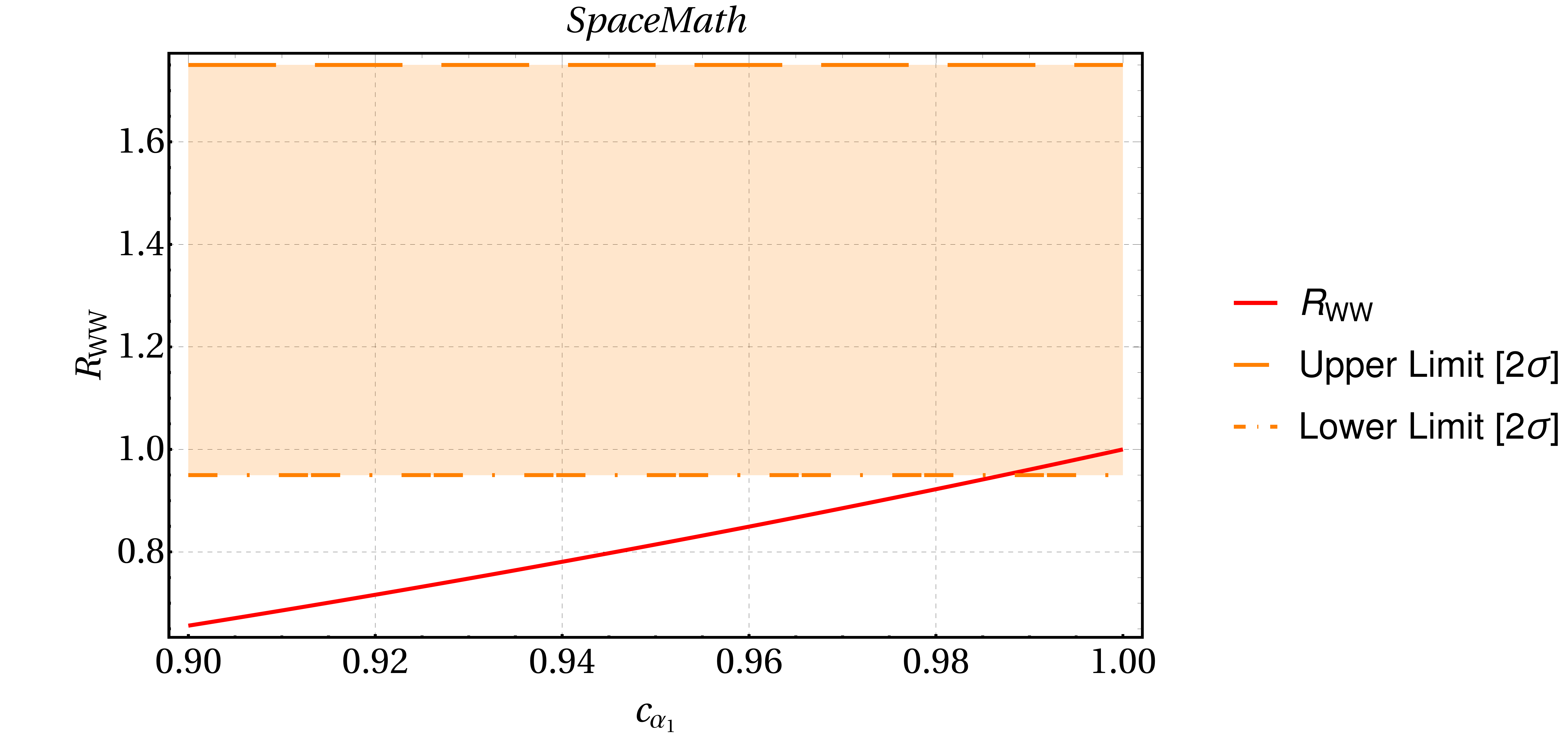}}
 \caption{$\mathcal{R}_{WW^*}$ as a function of $c_{\alpha_{1}}$. The dark area (orange online) represents the allowed region by the signal $\mathcal{R}_{WW^*}$ at $2\sigma$. \label{RWvsCalpha}   }
\end{figure}

Due to the coupling $g_{hPP}^{\text{IDMS}}=\cos{\alpha_{1}}\cdot g_{hPP}^{\text{SM}}$, with $P=f_i,W$, allowed regions for $\cos\alpha_1=c_{\alpha_{1}}$ can be extracted experimentally from $\mathcal{R}_{x\bar{x}}$. We find that $\mathcal{R}_{WW^*}$ is the most stringent way of limiting $c_{\alpha_{1}}$. In the fig. \ref{RWvsCalpha} we show the $c_{\alpha_1}-\mathcal{R}_{WW^*}$ plane, where the dark area (orange online) is the allowed region for $\mathcal{R}_{WW^*}$ at $2\sigma$.  The graph was generated via $\texttt{SpaceMath}$ \cite{SpaceMath}.
%
%
%
We note that the allowed interval for $c_{\alpha_{1}}$ is between $\sim 0.99-1$. This is to be expected since  $c_{\alpha_{1}}$ must be closed to the unit to have small deviations of the SM couplings. In particular, when $c_{\alpha_1}=1$ the SM is recovered. From now on we will consider $c_{\alpha_1}=0.99$.


\subsection{Constraint on the $Z^{\prime}$ gauge boson mass $m_{Z^{\prime}}$}
 In order to constrain the  $Z^{\prime}$ gauge boson mass, we now turn to analyze the $Z^{\prime}$ production cross-section times the branching ratio of $Z^{\prime}$ decaying into $\ell^-\ell^+$ ($\sigma_{Z^{\prime}}\mathcal{B}_{Z^{\prime}}$), with $\ell=e,\,\mu$. The ATLAS and CMS collaborations \cite{Aaboud:2017buh}, \cite{Sirunyan:2018exx}  searched for a new resonant and non-resonant high-mass phenomena in dilepton final states at $\sqrt{s}=13$ TeV with an integrated luminosity of 36.1 fb$^{-1}$ and 36 fb$^{-1}$, respectively. Nevertheless no significant deviation from the SM prediction was observed. Lower limits excluded on the resonant mass was reported depending on specific models. 
 
\begin{figure}[!h]
\center{\includegraphics[scale=0.35,angle=270]{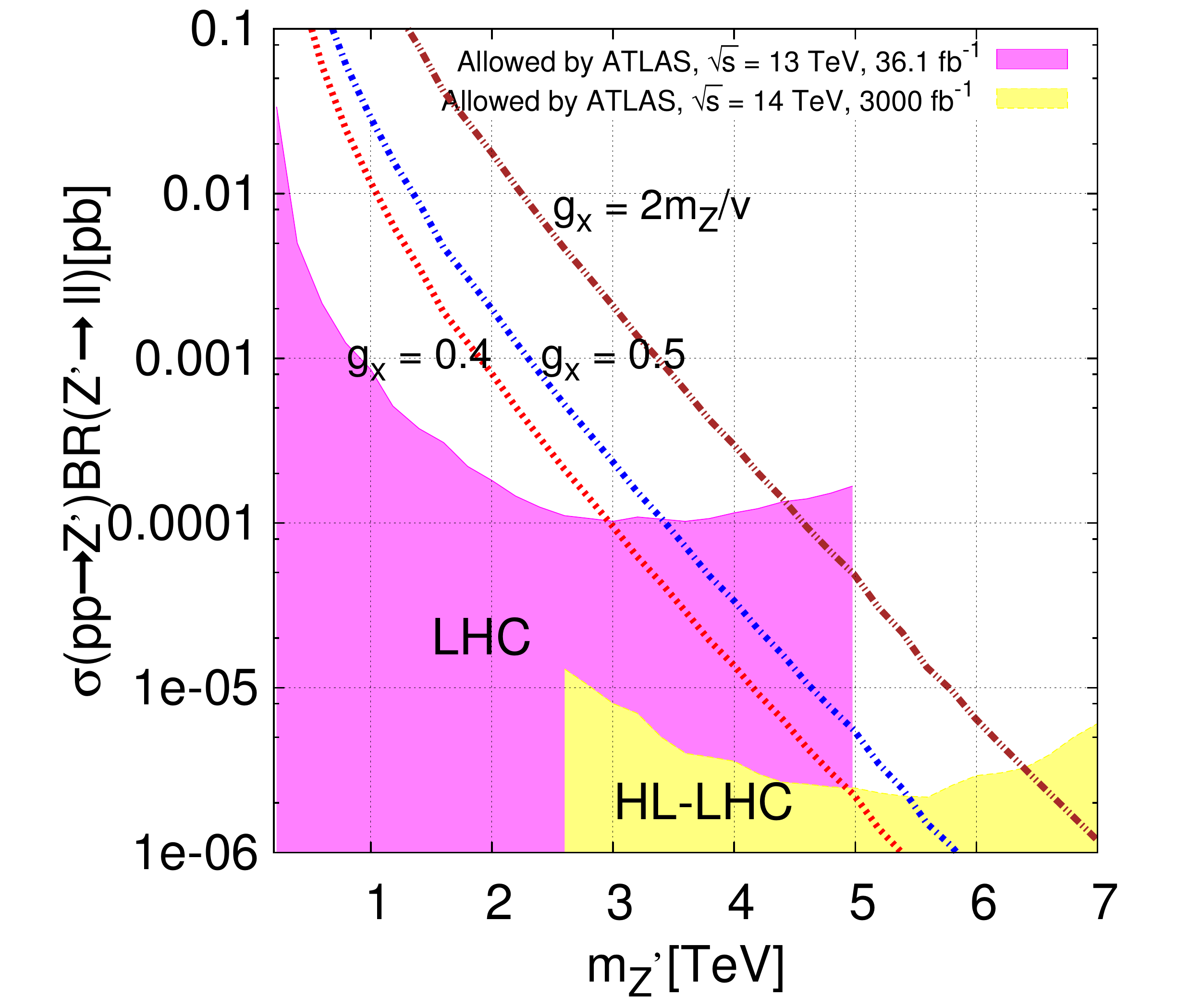}}
 \caption{$\sigma_{Z^{\prime}}\mathcal{B}_{Z^{\prime}}$ as a function of the $Z^{\prime}$ gauge boson mass for $g_x=0.4,\,0.5\,\text{and}\,2m_Z/\upsilon.$ Dark areas correspond to allowed regions by ATLAS collaboration \cite{Aaboud:2017buh}; magenta online corresponds to measurements at LHC and the yellow area represents a simulation for the HL-LHC \cite{ATL-PHYS-PUB-2018-044}. \label{Zpmass}  }
\end{figure}

Figure \ref{Zpmass} shows $\sigma_{Z^{\prime}}\mathcal{B}_{Z^{\prime}}$ as a function of the $Z^{\prime}$ gauge boson mass for $g_x=0.4,\,0.5$ and $2m_Z/\upsilon.$ The last value is related to the coupling of $Z$ gauge boson to fermions. We present two regions, the largest (magenta online) represents the results reported by ATLAS collaboration for a center-of-mass energy of $\sqrt{s}=13$ TeV and 36.1 fb$^{-1}$ as mentioned above, while the smallest (yellow online) area corresponds to $\sqrt{s}=14$ TeV and 3000 fb$^{-1}$, which is the goal of the High Luminosity Large Hadron Collider \cite{ATL-PHYS-PUB-2018-044}; these analyses are based on generator-level information with parameterized estimates applied to the final state particles to simulate the response of the upgraded ATLAS detector and pile-up collisions.

%
%
%
%
%
%
Considering the results reported by LHC (HL-LHC),\\ $\sigma_{Z^{\prime}}\mathcal{B}_{Z^{\prime}}>10^{-4}$ pb ($\sim 10^{-6}$ pb) excludes $m_{Z^{\prime}}\lesssim 3$ TeV ($m_{Z^{\prime}}\lesssim 5$ TeV) for $g_x=0.4$, while $m_{Z^\prime}\lesssim 3.4$ TeV ($m_{Z^{\prime}}\lesssim 5.4$ TeV) for $g_x=0.5$ are excluded. Finally, we explored the case in which $g_x=g_Z=2m_Z/\upsilon$ and we observe a similar behavior as reported in the refs. \cite{Aaboud:2017buh}-\cite{ATL-PHYS-PUB-2018-044}, excluding $m_{Z^{\prime}}\lesssim 4.5$ TeV ($m_{Z^{\prime}}\lesssim 6.5$ TeV) . 


\subsection{Constraint on $\upsilon_x$, $g_x$}
\begin{figure}[!h]
\center{\includegraphics[scale=0.2]{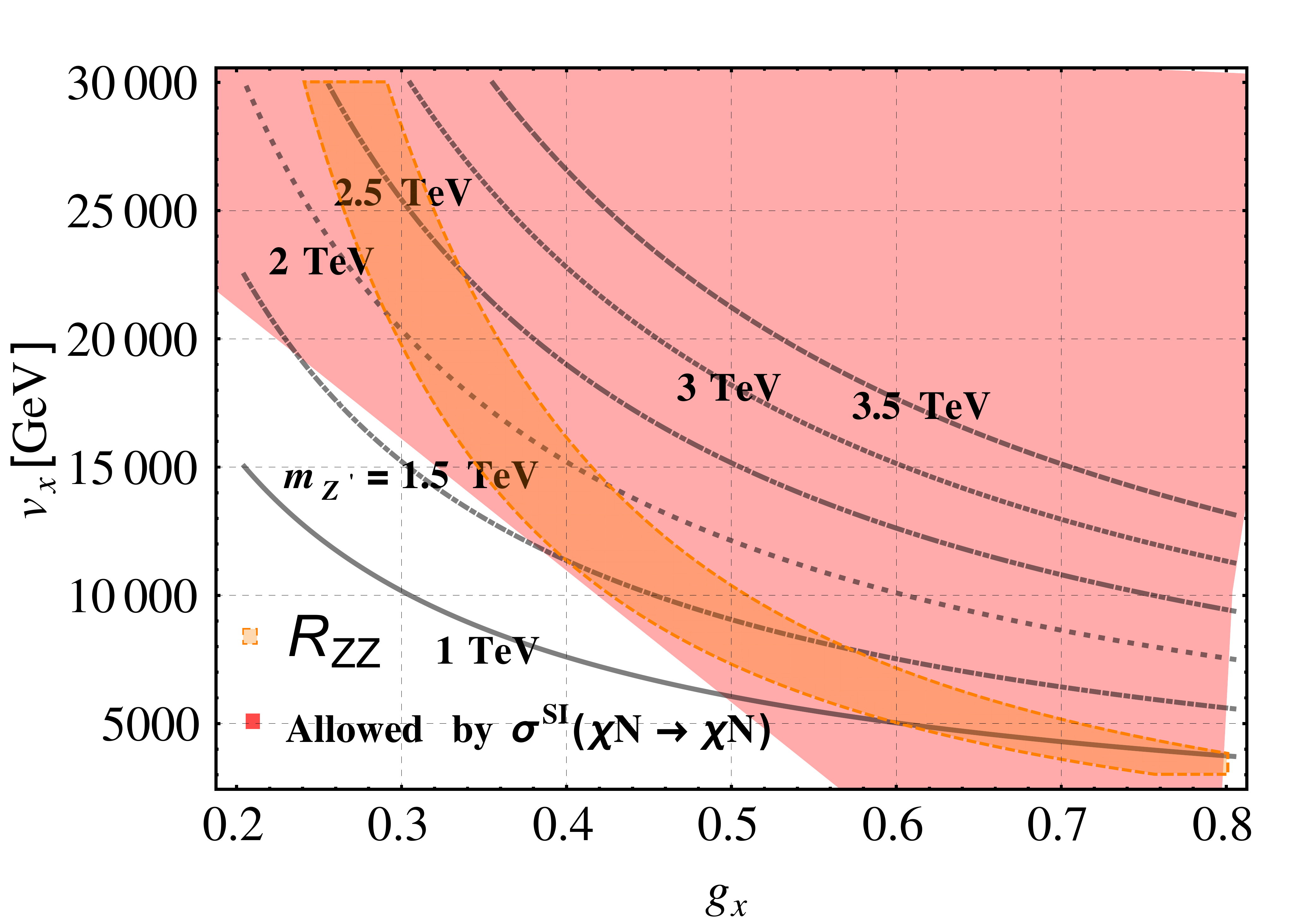}}
 \caption{The inclined and curved dark area (orange online) represents the consistent region with $R_{ZZ^*}$ while the large area (light red) indicates the allowed values by the upper limit on $\sigma^{SI}(\chi N\to \chi N)$. The white area represents the excluded region by both observables. Curved lines represent the predicted value of $m_{Z^{\prime}}$ as a function of $\upsilon_x$ and $g_x$.  \label{Vx-gx}   }
\end{figure}

In the fig. \ref{Vx-gx} we show the $g_x-\upsilon_x$ plane, in which allowed regions for $\mathcal{R}_{ZZ^*}$ and the upper limit on WIMP-nucleon cross section, $\sigma^{SI}(\chi N\to \chi N)$, are displayed. We generate the Feynman rules of the IDMS via $\texttt{LanHEP}$ \cite{lanhep} and we evaluate $\sigma^{SI}(\chi N\to \chi N)$ through $\texttt{CalcHep}$ \cite{calchep}. We observe that the consistent zone with both $\mathcal{R}_{ZZ^*}$ and $\sigma^{SI}(\chi N\to \chi N)$ allows values for $\upsilon_x$ in the interval from $\sim 11$ to $\sim 16$ TeV for $g_x=0.4$, while the white area represents the excluded region.
\subsection{Constraint on the charged scalar boson mass $m_{H^{\pm}}$}
We use $\mathcal{R}_{\gamma\gamma}$ in order to constrain the charged scalar boson mass $m_{H^{\pm}}$. In addition to the SM contributions, the $h\to\gamma\gamma$ decay receives contributions at one-loop level of charged scalar bosons predicted by the IDMS. The $b\to s\gamma$ decay is another process that can also impose strong restrictions on the charged scalar boson mass. However, since this particle arises from the inert doublet, its Yukawa couplings with fermions are absent, so the $b\to s \gamma$ decay is not a way to restrict the charged scalar boson mass.

\begin{figure}[!h]
\center{\includegraphics[scale=0.6]{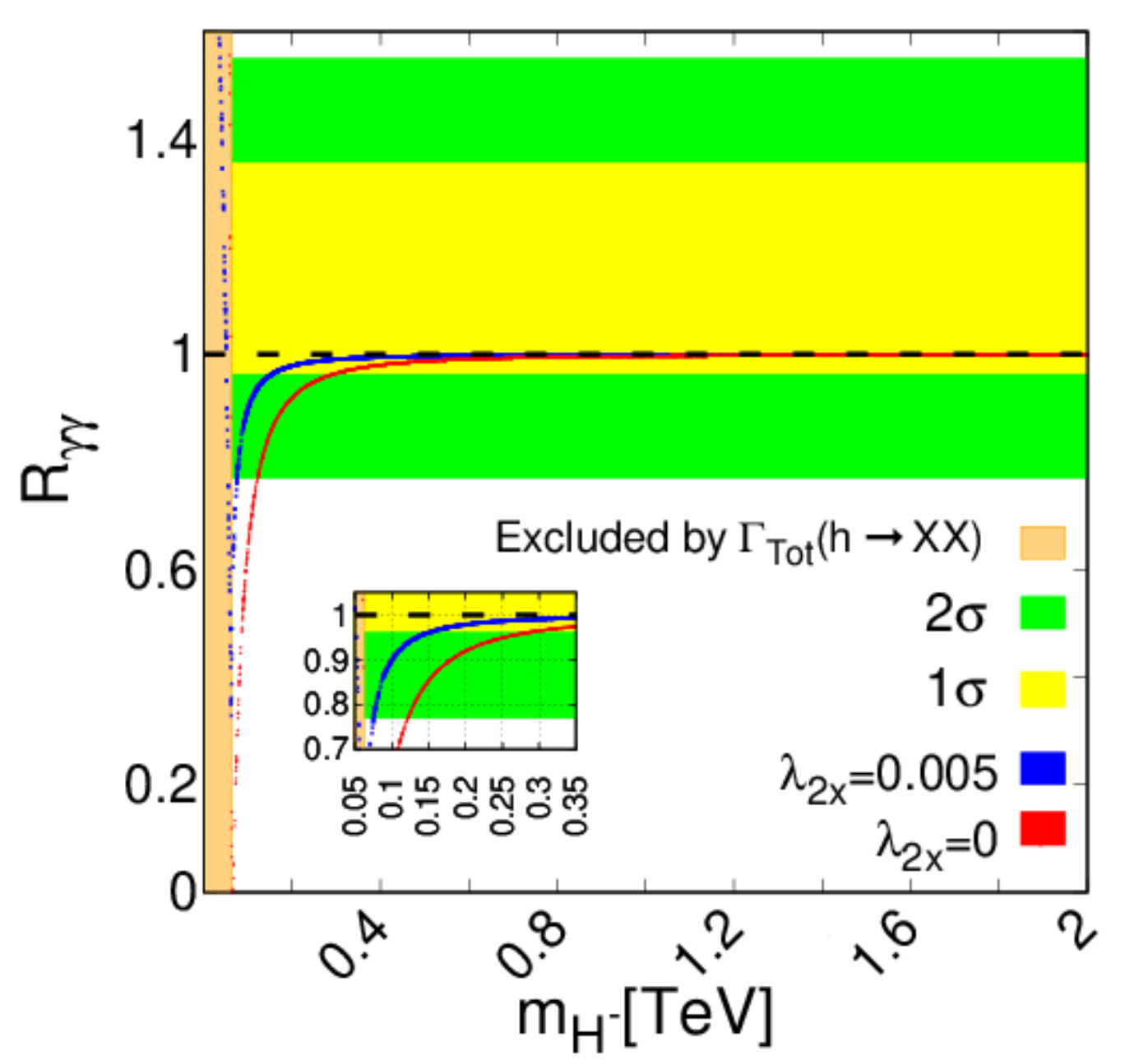}}
 \caption{Diphoton rate as a function of the charged scalar boson mass, $m_{H^{\pm}}$. Values of $\mathcal{R}_{\gamma\gamma}$ allowed at 1$\sigma$ and 2$\sigma$ are represented by the thin (yellow online)  and broad (green online) horizontal bands, respectively. While vertical band (orange online) is the excluded region for $m_{H^{\pm}}$ by total decay width of the Higgs boson. \label{Vx-mch}   }
\end{figure}

Figure \ref{Vx-mch} shows the $m_{H^{\pm}}-\mathcal{R}_{\gamma\gamma}$ plane. We note that $\mathcal{R}_{\gamma\gamma}$ imposes a lower bound on the charged scalar boson mass as $330$ GeV  $\lesssim m_{H^{\pm}}$ at 1$\sigma$ with $\lambda_{2x}=0$. Figure \ref{Vx-mch} also shows bounds for $\lambda_{2x}=0.005$, which are less restrictive than the previous case. We find that $170$ GeV $\lesssim m_{H^{\pm}}$ ($75$ GeV $\lesssim m_{H^{\pm}}$) at 1$\sigma$ (2$\sigma$), respectively. Finally, the total decay width of the Higgs boson \cite{Tanabashi:2018oca} excludes $60$ GeV $\lesssim m_{H^{\pm}}$. The values for $\lambda_{2x}$ parameter are select such that they are compatibles with viable values for relic density within the framework of the IDMS. We generated random values for $\lambda_3$ between $0.01-0.0105$ ($0.0297-0.03$) for $\lambda_{2x}=0.005$ ($\lambda_{2x}=0$), respectively, for the same reason.

In table \ref{tableparameters} we present a summary of the values for the model parameters used in our following analysis.

\begin{table}
\caption{The setting of the values for the model parameters. }
\begin{centering}\label{tableparameters}
\begin{tabular}{ccc}
\hline
Parameter & Value & Constraint\tabularnewline
\hline
\hline
$c_{\alpha_{1}}$ & $0.99$ & $\mathcal{R}_{WW}$\tabularnewline
\hline
$g_{x}$ & $0.4$ & $\mathcal{R}_{ZZ}$ and $\sigma^{SI}(\chi N\to \chi N)$\tabularnewline
\hline
$\upsilon_{x}$ & $23$ TeV&  $\mathcal{R}_{ZZ}$ and $\sigma^{SI}(\chi N\to \chi N)$\tabularnewline
\hline
$m_{H^{\pm}}$ & $0.5$ TeV &  $\mathcal{R}_{\gamma\gamma}$\tabularnewline
\hline
$m_{Z^{\prime}}$ & $3$ TeV&  $\sigma(pp\to Z^{\prime})\mathcal{B}(Z^{\prime}\to \ell\ell)$ \tabularnewline
\hline
\end{tabular}
\par\end{centering}
\end{table}

%
\section{Phenomenology for $S$ and $Z^{\prime}$}
\label{sec4}

\begin{figure}[htb!]
\centering
\includegraphics[scale=0.4,angle=270]{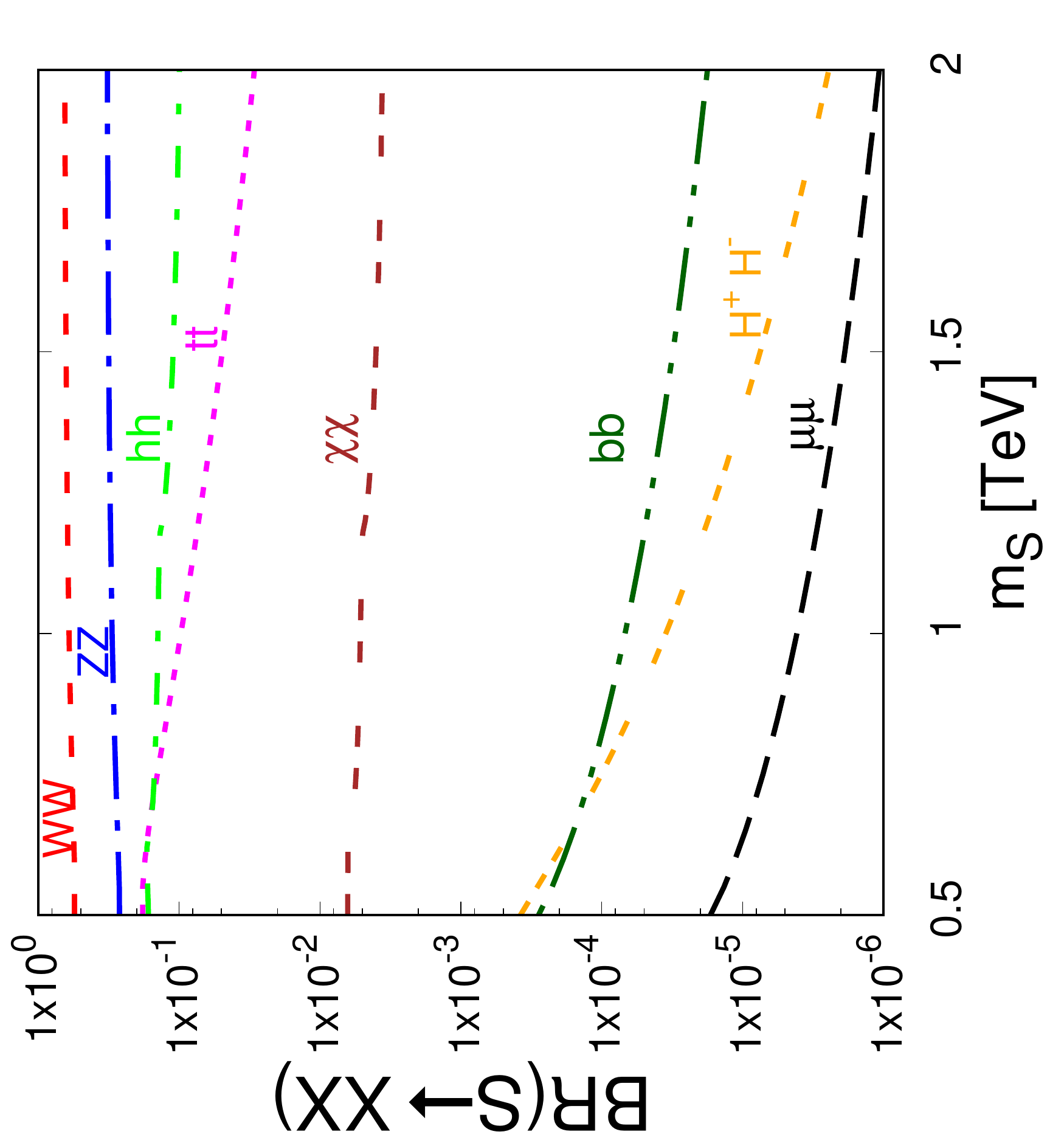}
\includegraphics[scale=0.4,angle=270]{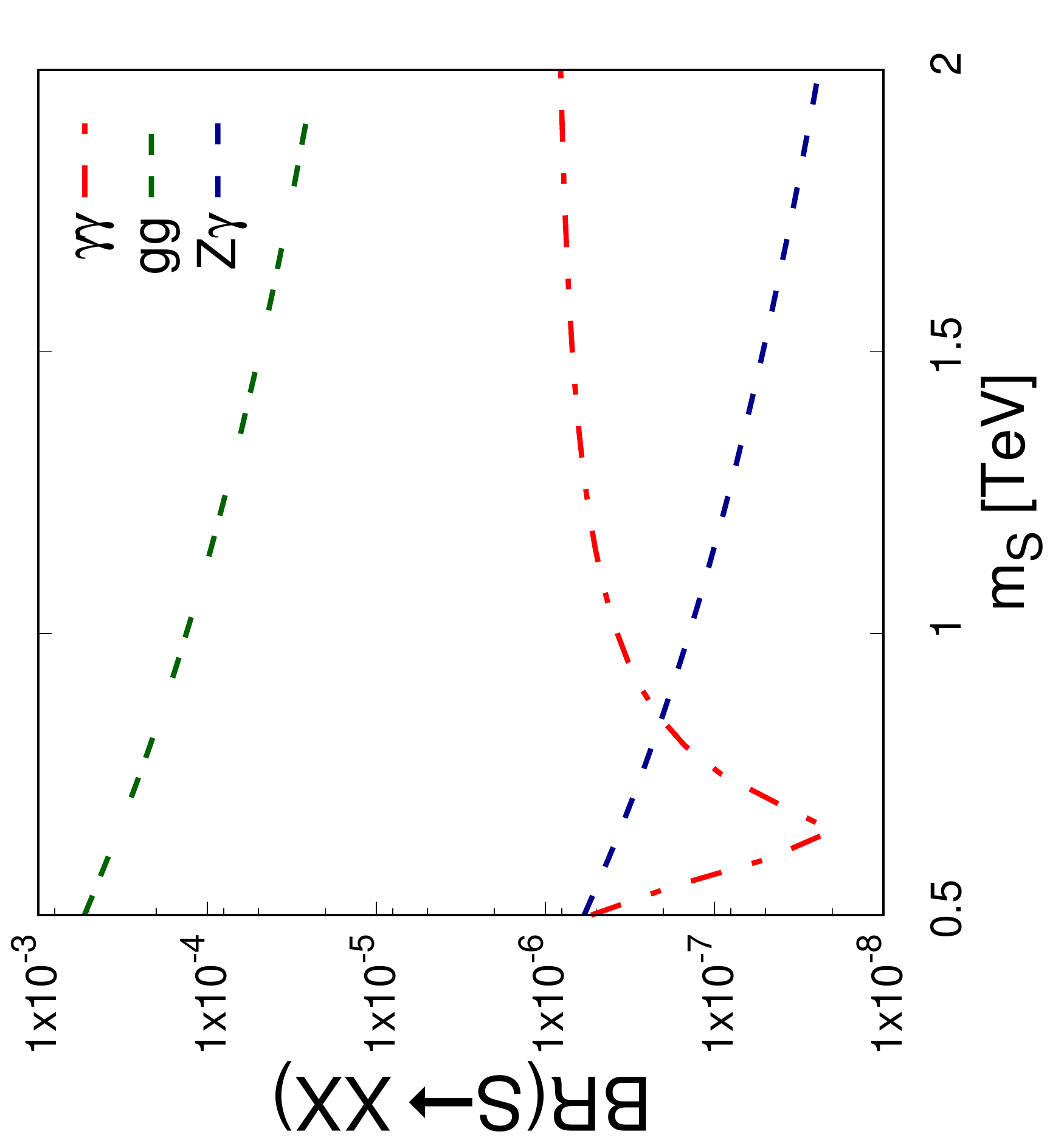}
\caption{Branching ratios of scalar $S$ as a function of its mass. Top: tree-level decays; bottom: one-loop level decays.
\label{BR_S-XX}}
\end{figure}

\begin{figure}[htb!]
\centering
\includegraphics[scale=0.4,angle=270]{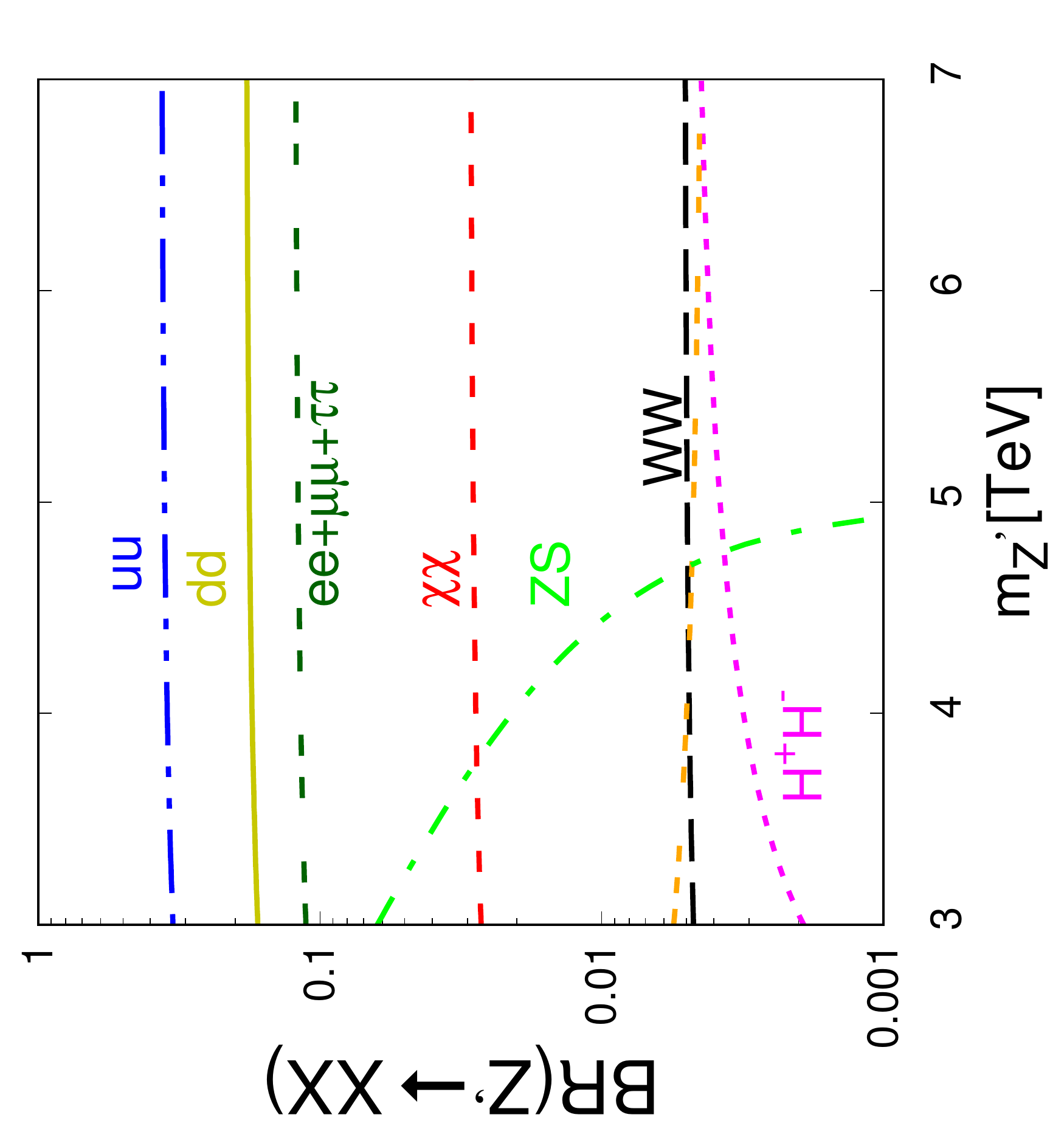}
\caption{Relevant branching ratios of $Z^{\prime}$ gauge boson as a function of its mass. 
\label{BR_Zp-XX}}
\end{figure}

We now analyze the behavior of the branching ratio of the dominant decay channels of the particles coming from complex singlet, namely, scalar $S$ and the $Z^{\prime}$ gauge boson. Analytical formulas of the partial decay widths are presented in \ref{App-A}. 
Figure \ref{BR_S-XX} shows the branching ratios for relevant decays of the $S$ neutral scalar at tree and one-loop level. In Figure \ref{BR_Zp-XX} the relevant decay channels are also presented but for the $Z^{\prime}$ gauge boson predicted by the IDMS.

We observe that the dominant $S$ decay modes are $S\to VV$, with $V=W,\,Z$, and $S\to hh$. These processes are of the order of $10^{-1}$. Once the $t\bar{t}$ channel became open for $m_S\geq 2m_t$, its branching ratio is of the order of $S\to VV$ channel up to a $S$ mass of about $800$ GeV. Later when $m_S$ increases, the value of the $BR(S\to t\bar{t})$  decreases such that $BR(S\to t\bar{t})\sim \mathcal{O}(10^{-2})$ for $m_S=2$ TeV. Another relevant decay mode is $S\to b\bar{b}$ whose branching ratio range decreases from $10^{-3}$ to $10^{-5}$. At one-loop level, the dominant channel is $S\to gg$ with a branching ratio up to $\mathcal{O}(10^{-4})$ for $m_S=200$ GeV and $\mathcal{O}(10^{-5})$ for $m_S=2$ TeV.

 As far as the $Z^{\prime}$ gauge boson is concerned, their dominant decay modes are into type-up quarks, whose sum is about $3.5\times 10^{-1}$, followed by type-down quarks and finally by charged leptons with a branching ratio of the order $10^{-1}$.

\section{Relic density}
\label{sec5}
Once the model parameters are bounded by experimental and theoretical constraints, we now turn to analyze if the model can help us to understand the relic density, which is the current experimental quantity of DM particle that remains after of freeze-out process. The observed value for non-baryonic matter reported by PLANCK collaboration \cite{PLANCK} is
\begin{equation}
\Omega h^2 = 0.120\pm0.001,
\end{equation}
where $h$ is the Hubble constant in units of $100 \frac{km}{s. Mpc}$. Our analysis is based on selecting $\chi$ as DM candidate. The relic density is obtained by solving the Boltzmann equation for the number density rate which is given by
\begin{equation}
a^{-3} \frac{d}{dt}(na^3) = \langle\sigma v\rangle  (n^2_{eq} - n^2),
\end{equation}
where $n$ is the DM number density and $a$ is a scale factor. All information about the model is contained in the thermally averaged cross section $\langle\sigma v\rangle$.
The relic density, $\Omega h^2$, is obtained by using the \texttt{micrOmegas} package \cite{micromegas,micromegas1}, which requires all information about the IDMS for which we implement the model via the \texttt{LanHep} package \cite{lanhep}. 

\begin{figure}[htb!]
\centering
\includegraphics[scale=0.5,angle=270]{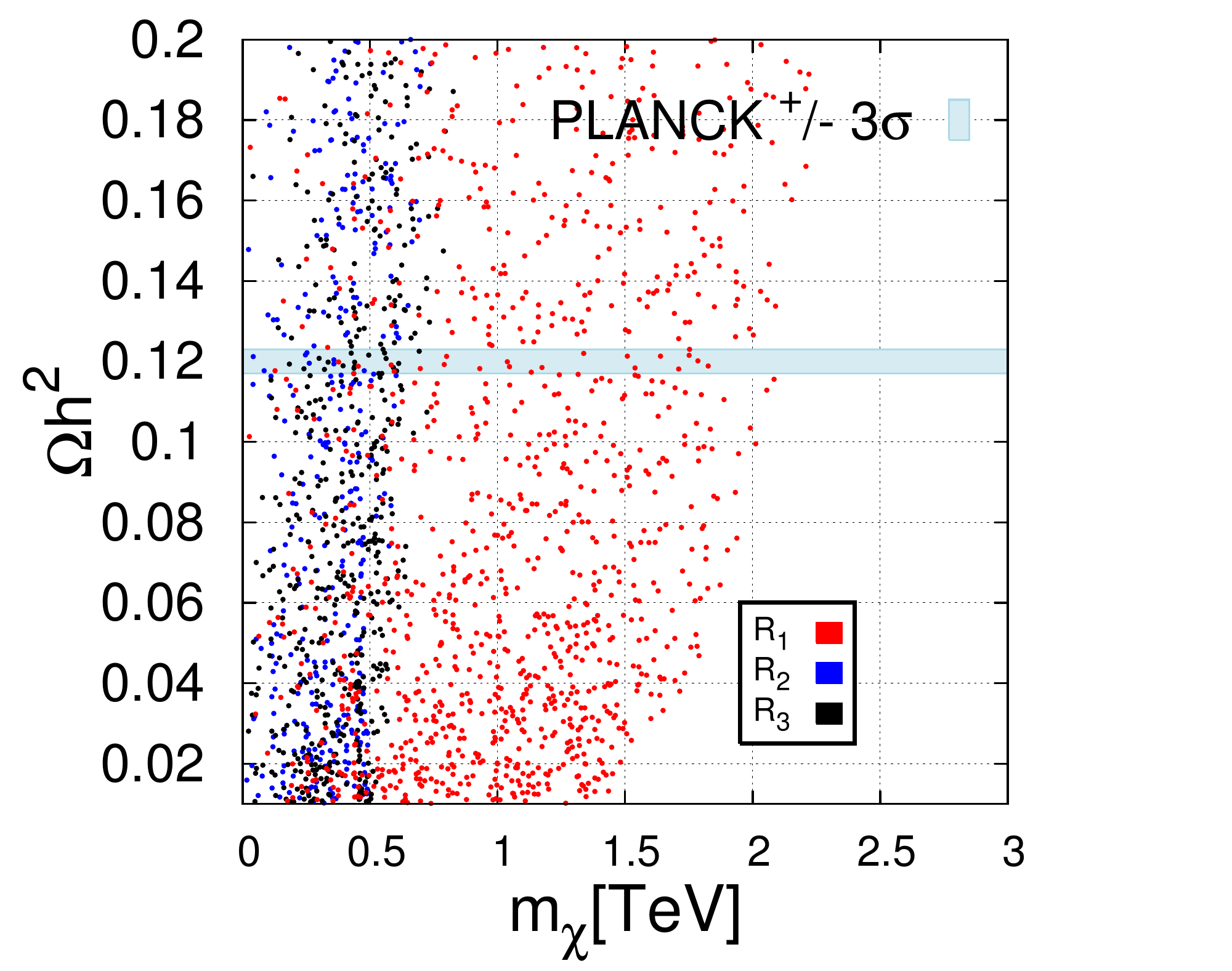}
\caption{Relic density as a function of the DM candidate mass. $R_i$ scenarios are described in the main text.
\label{GlobalRelicDensity}}
\end{figure}

In Fig. \ref{GlobalRelicDensity} we present a scattering plot of the relic density as a function of the DM candidate mass, $m_{\chi}$. We show three scenarios to note the sensitivity of the relic density on $\lambda_{2x}$. These scenarios are classified by their random value intervals:
\begin{itemize}
\item $R_1:\lambda_{2x}\sim\mathcal{O}(10^{-3}-10^{-2})$,
\item $R_2:\lambda_{2x}\sim\mathcal{O}(10^{-4}-10^{-3})$,
\item $R_3:\lambda_{2x}= 0 \cup \mathcal{O}(10^{-7}-10^{-3})$.
\end{itemize}
In all cases, we use  random values for $\lambda_{345}\sim\mathcal{O}(10^{-2}-10^{-1})$ and values for $m_{\chi}$ from $1$ to $3000$ GeV. 
The most favored scenario is $R_3$, which contains the special case $\lambda_{2x}=0$. When this occurs, the IDM $h\chi\chi$ coupling is recovered, as the Table \ref{FR-IDMS} shows. Nevertheless, in the IDMS two new portals contribute to relic density, namely, $Z^{\prime}$ gauge boson and the neutral scalar $S$; both arise from the complex singlet $\mathcal{S}_X$. We observe that, depending on $\lambda_{345}$ and $\lambda_{2x}$, masses from a few GeV to about 2 TeV are in agreement with the results for the relic density of the PLANCK collaboration \cite{PLANCK}. It is worth mentioning that we include a pair of photons in the final state in the process of annihilation of the DM particles, i.e., $\chi\chi\to h\to \gamma\gamma$.

\begin{figure}[htb!]
\centering
\includegraphics[scale=0.35,angle=270]{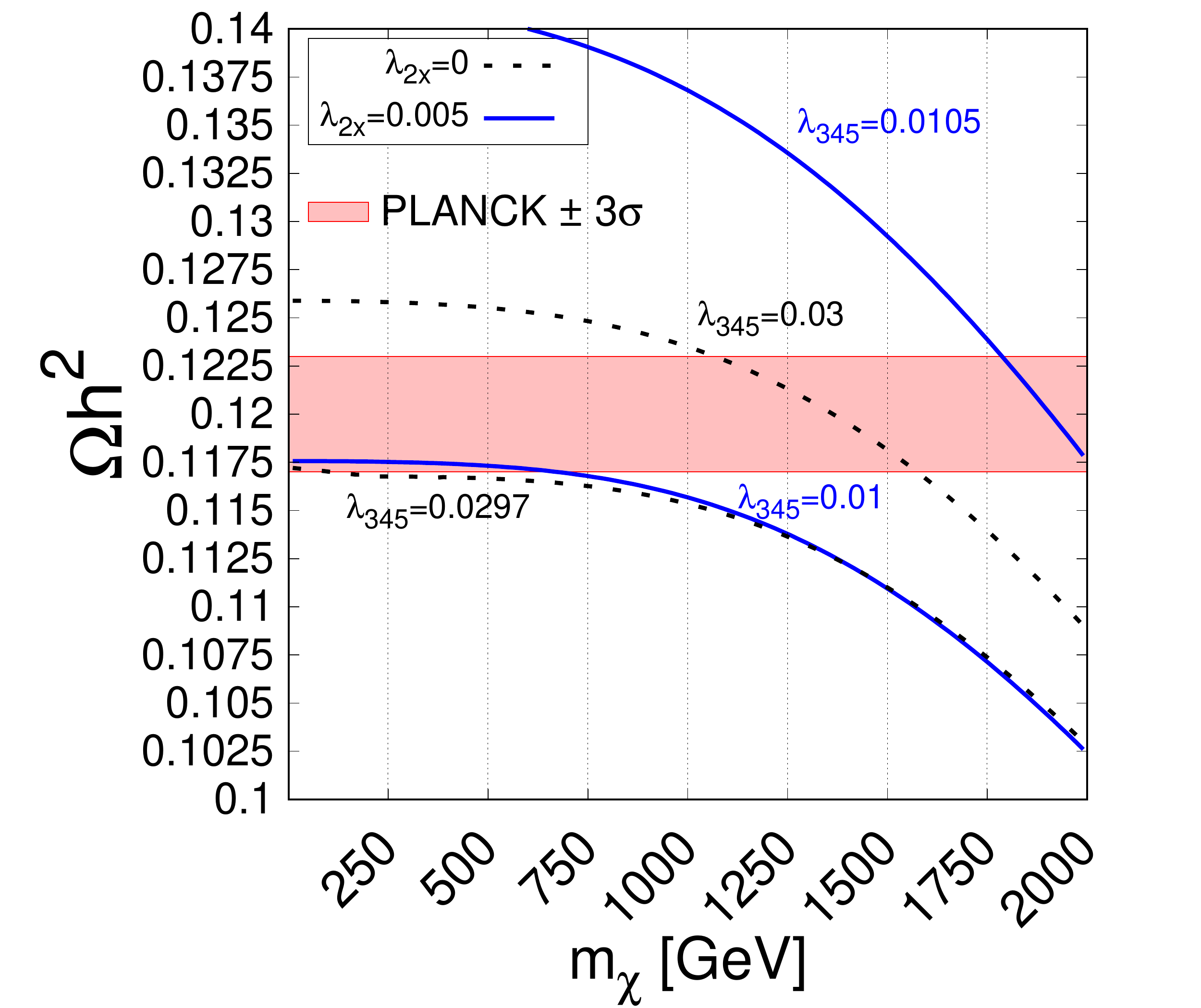}
\caption{Relic density as a function of the DM candidate mass for $\lambda_{345}=0.0297,\,0.03$ ($0.01,\,0.0105$) with $\lambda_{2x}=0$ ($\lambda_{2x}=0.005$).
\label{RelicDensity}}
\end{figure}

Figure \ref{RelicDensity} shows the representatives values of $\lambda_{345}$ in the interval $0.0297-0.03$ ($0.01-0.0105$) and $\lambda_{2x}=0$ ($\lambda_{2x}=0.005$), respectively. The scattering process $\sigma^{SI}(\chi N\to \chi N)$ excludes an important region of allowed values of $g_x$ and $\upsilon_x$ for $\lambda_{345}>0.0105$ ($\lambda_{345}>0.03$) by assuming $\lambda_{2x}=0$ ($\lambda_{2x}=0.005$). Under these considerations we find intervals for the masses of DM candidates:
\begin{enumerate}
\item For $\lambda_{2x}=0:$
\begin{itemize}
\item Light masses: $1\lesssim m_{\chi}\lesssim80$ GeV.
\item Heavy masses: $1100\lesssim m_{\chi}\lesssim1600$ GeV.
\end{itemize}
\item For $\lambda_{2x}=0.005:$
\begin{itemize}
\item Light and intermediate masses: $1\lesssim m_{\chi}\lesssim700$ GeV.
\item Heavy masses: $1800\lesssim m_{\chi}\lesssim2100$ GeV.
\end{itemize}
\end{enumerate}


\section{Dark Matter production at hadron colliders}
\label{sec6}
The ATLAS and CMS collaborations \cite{Aaboud:2017dor} searched for the reaction $pp\to \chi\chi\gamma$ through events that contain an energetic photon and large missing transverse momentum, corresponding to an integrated luminosity of 36.1 fb$^{-1}$ at centre-of-mass energy of 13 TeV. However, only the exclusion limits were reported. In order to motivate a potential and sophisticated study of the production of DM particles at hadron colliders, we evaluate the $pp\to \chi\chi\gamma$ production cross section and the main SM background processes via $\texttt{MadGraph5}$ \cite{Alwall:2011uj}. Our study is focused on future hadron colliders, namely:
\begin{itemize}
\item High-Luminosity Large Hadron Collider \cite{Apollinari:2017cqg} (HL-LHC). The HL-LHC is a new stage of the LHC starting about 2026 with a center-of-mass energy of 14 TeV. The upgrade aims at increasing the integrated luminosity by a factor of ten ($\sim$3000 fb$^{-1}$) with respect to the final stage of the LHC ($300$ fb$^{-1}$).
\item High-Energy Large Hadron Collider \cite{Benedikt:2018ofy} (HE-LHC). The HE-LHC is a possible future project at CERN. The HE-LHC will be a 27 TeV $pp$ collider being developed for the 100 TeV Future Circular Collider. This project is designed to reach up to 12000 fb$^{-1}$ which opens a large window for new physics research.
\item Future Circular hadron-hadron Collider \cite{Arkani-Hamed:2015vfh} (FCC-hh). The FCC-hh is a future 100 TeV $pp$ hadron collider which will be able to discover rare processes, new interactions up to masses of around 30 TeV and search for a possible substructure of the quarks. The FCC-hh will reach up to an integrated luminosity of 30000 fb$^{-1}$ in its final stage.
\end{itemize}

\subsection{Signal and background events}
\label{SGN-BGD}
The main SM background to the $\gamma+E_T^{miss}$ final state are events containing either a true photon or an object misidentified as a photon. The dominant background processes are the electroweak production of $Z(\to\nu\nu)\gamma$, $W(\to\ell\nu)\gamma$ and $Z(\to\ell\ell)\gamma$ with unidentified charged leptons, $e,\,\mu$, or with $\tau\to$hadrons+$\nu_{\tau}$. \\

As far as our computation scheme is concerned, we first use the $\texttt{LanHEP}$ \cite{lanhep} routines to obtain the IDMS Feynman rules for $\texttt{MadGraph5}$ \cite{Alwall:2011uj}. Secondly, we evaluated the production cross section of the signal and background processes ($\textbf{PCSS}$ and $\textbf{PCSB}$) and we generated $10^{5}$ events for both reactions. 

\begin{figure}[!htb]
\centering
 \subfigure[ ]{\includegraphics[scale=0.35,angle=270]{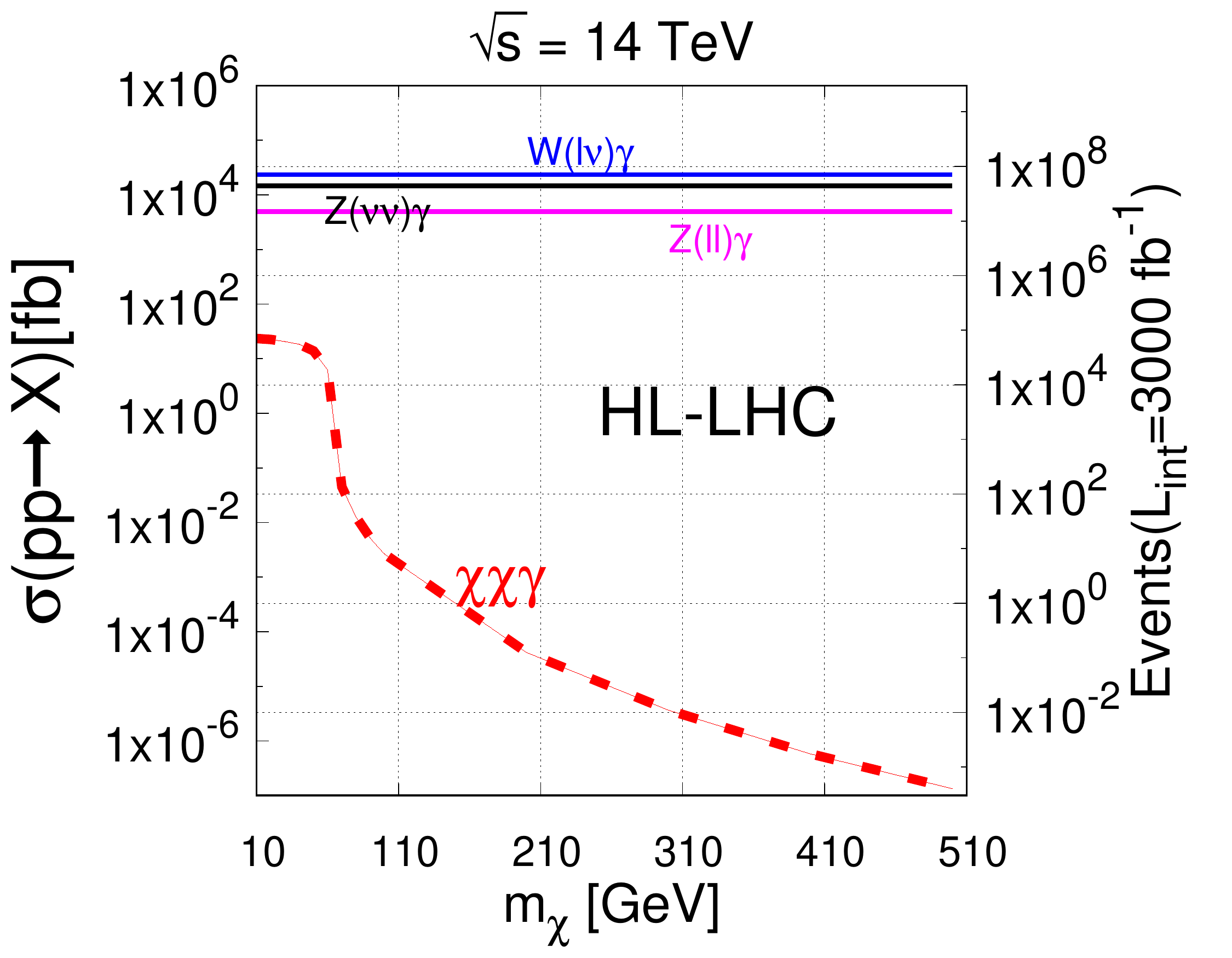}\label{Events14TeV}}
 \subfigure[ ]{\includegraphics[scale=0.35,angle=270]{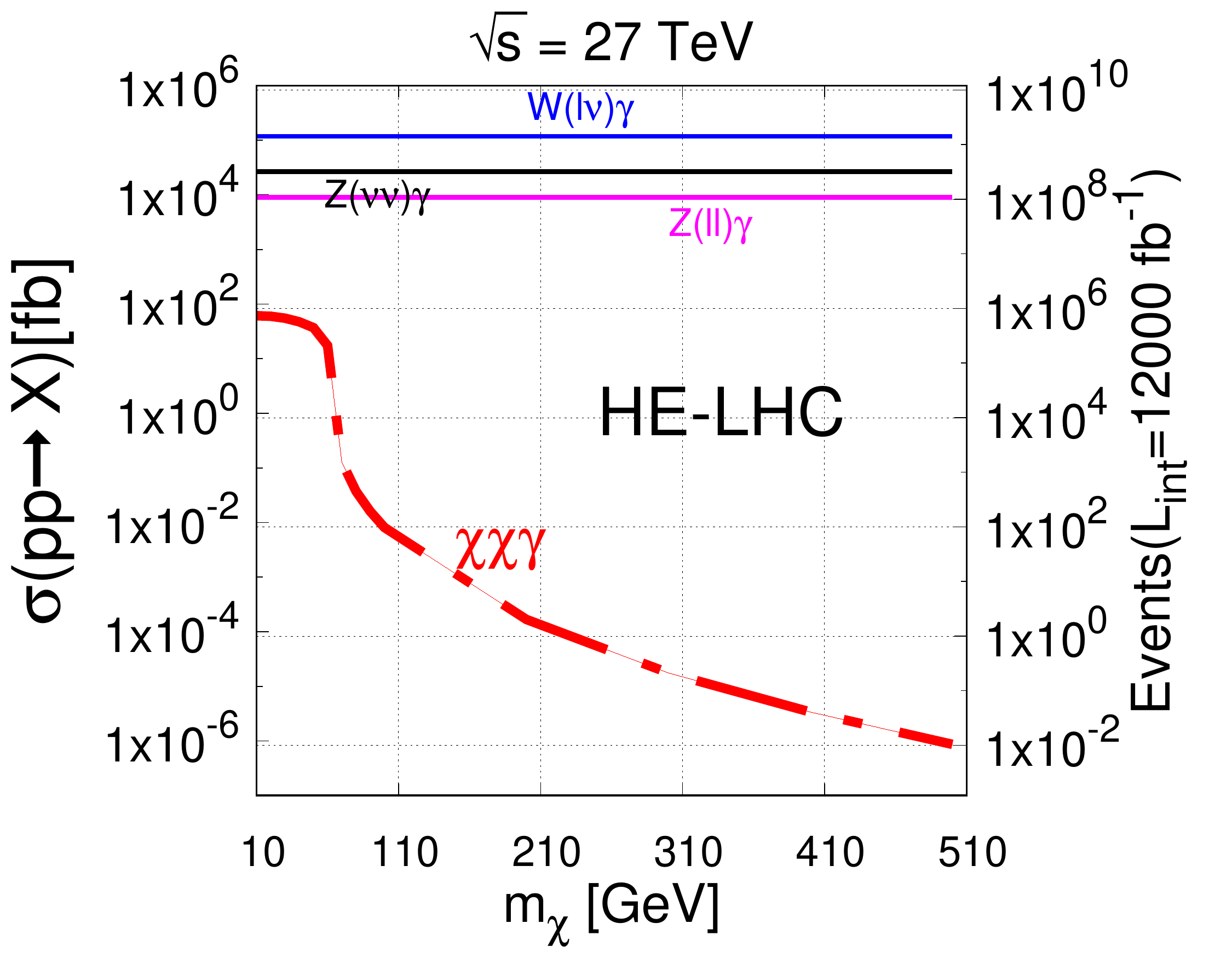}\label{Events27TeV}}
 \subfigure[ ]{\includegraphics[scale=0.35,angle=270]{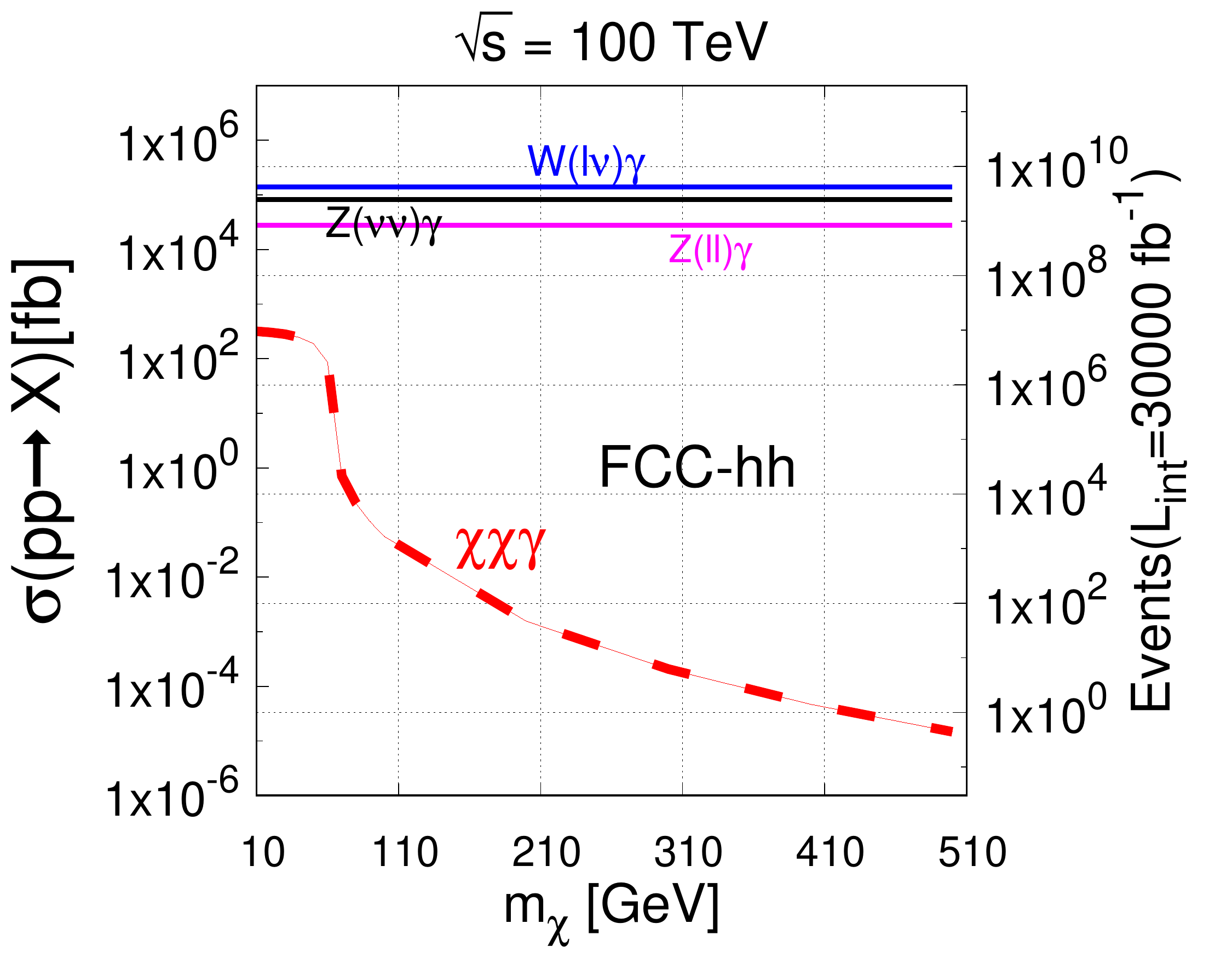}\label{Events100TeV}}
 \caption{On the left axis: production cross section for the signal $pp\to\chi\chi\gamma$ and SM background processes $W(\to\ell\nu)\gamma$, $Z(\to\nu\nu)\gamma$, $Z(\to\ell\ell)\gamma$. On the right axis: number of events produced. (a) \textbf{HL-LHC} at $\sqrt{s}=14$ TeV and $\mathcal{L}_{int}=3$ ab $^{-1}$, (b) \textbf{HE-LHC} at $\sqrt{s}=27$ TeV and $\mathcal{L}_{int}=12$ ab $^{-1}$, (c) \textbf{FCC-hh} at $\sqrt{s}=100$ TeV and $\mathcal{L}_{int}=30$ ab $^{-1}$ .   \label{Events}}
	\end{figure}

In fig. \ref{Events}, we present the $\textbf{PCSS}$ and $\textbf{PCSB}$ (axis left) and number of events (axis right) for the different future hadron colliders, i.e.,  \ref{Events14TeV} HL-LHC, \ref{Events27TeV}  HE-LHC and finally \ref{Events100TeV} for the FCC-hh, with integrated luminosities 3000 fb$^{-1}$, 12000 fb$^{-1}$, 30000 fb$^{-1}$, respectively. In all graphics, horizontal lines represent the potential SM background processes. We observe that light masses for the DM candidate are favored producing up to about $10^{5}$ ($10^{6}$, $10^{7}$) events at the HL-LHC (HE-LHC, FCC-hh) by considering a DM mass of 10 GeV. However, the intermediate regimen of masses ($\sim$ 500 GeV) is disadvantaged by this channel, even at the FCC-hh only one event will be produced. Therefore, we analyze the range of masses 10-100 GeV for event reconstruction. 

\subsection{Event reconstruction}
\begin{figure}[!htb]
\centering
 \subfigure[ ]{\includegraphics[scale=0.4]{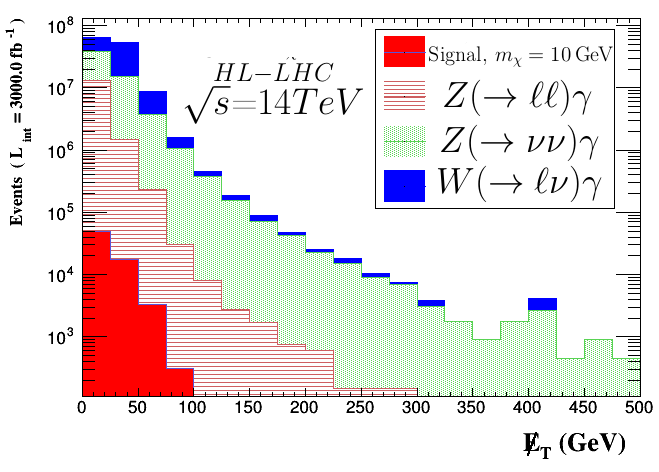}}
 \subfigure[ ]{\includegraphics[scale=0.4]{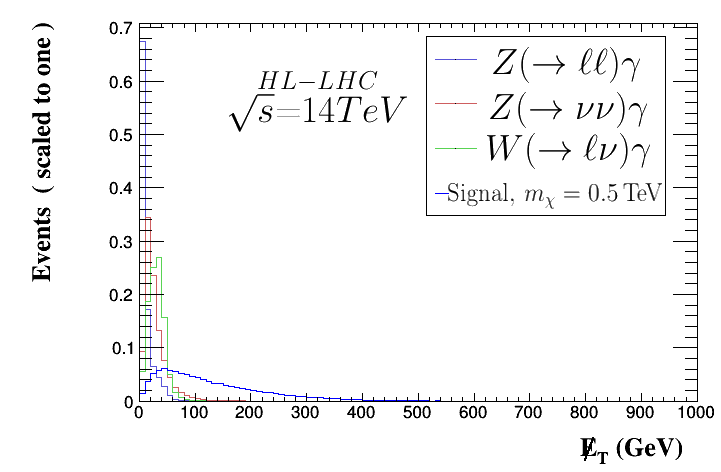}}
 \caption{Distribution of $E_T^{\text{miss}}$ with no cuts for signal and main background processes. (a) $m_{\chi}=10$ GeV; (b) $m_{\chi}=0.5$ TeV and normalized to one.    \label{MET}}
	\end{figure}

\begin{figure}[!htb]
\centering
 \subfigure[ ]{\includegraphics[scale=0.4]{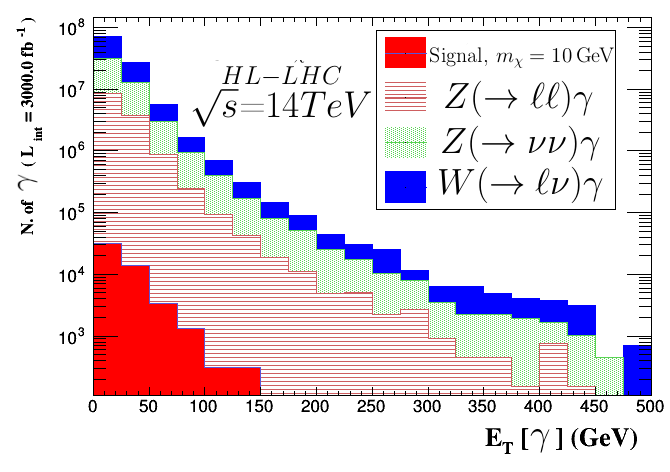}}
 \subfigure[ ]{\includegraphics[scale=0.4]{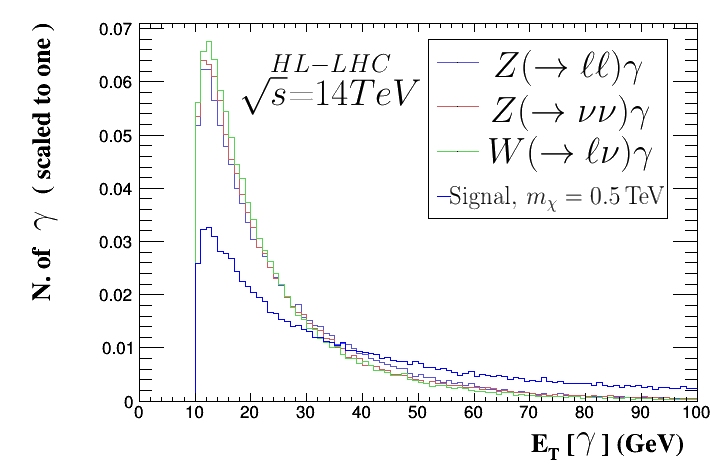}}
 \caption{Distribution of $E_T^{\gamma}$ with no cuts for signal and main background processes. (a) $m_{\chi}=10$ GeV; (b) $m_{\chi}=0.5$ TeV and normalized to one. \label{ET}}
	\end{figure}

\begin{figure}[!htb]
\centering
 \subfigure[ ]{\includegraphics[scale=0.4]{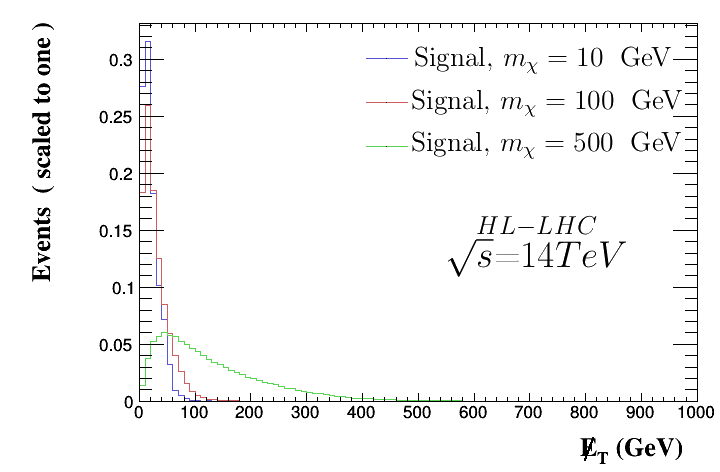}}
 \subfigure[ ]{\includegraphics[scale=0.4]{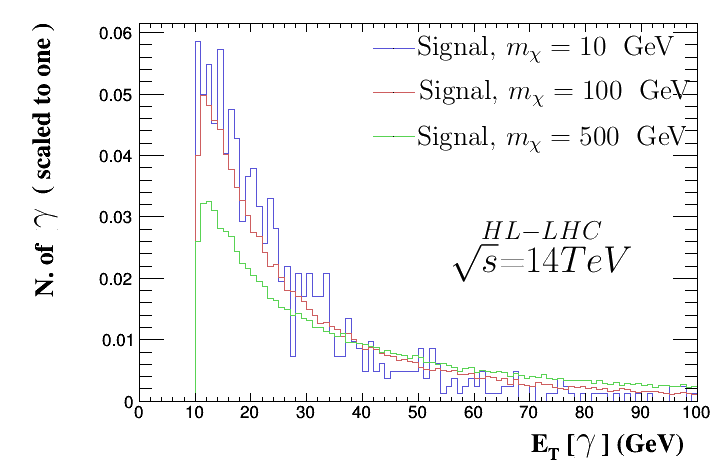}}
 \caption{(a) Distribution of $E_T^{\text{miss}}$ and (b) distribution of $E_T^{\gamma}$ for $m_{\chi}=$10, 100 and 500 GeV. \label{DMmasses-MET-ET}}
	\end{figure}

We closely follow the strategy by ATLAS Collaboration \cite{Aaboud:2017dor} in which the photon identification is based on energy deposited at the electromagnetic calorimeter. Candidate photons are required to have $E_T^{\gamma}>150$ GeV, to be within $|\eta|<1.37$ and be isolated by demanding energy in the calorimeter in a cone of size $\Delta R=\sqrt{(\Delta\eta)^2+(\Delta\phi)^2}=0.4$. Due to the elusive nature of the DM candidates, these particles are characterized by missed energy transverse and therefore we demand for $E_T^{\text{miss}}$> 150 GeV. It is also required that the photon and the $E_T^{\text{miss}}$ do not overlap in the azimuthal plane, then is required the condition $\Delta\phi(\gamma,\,E_T^{\text{miss}})>0.4$. In our analysis, the above requirements work well for intermediate masses, 100 GeV $\lesssim m_{\chi}$. In the analysis performed in Section \ref{SGN-BGD}, masses in the interval of 10-100 GeV (10-300 GeV), for HL-LHC and HE-LHC (for FCC-hh), respectively, are favored. Therefore, for light DM masses we apply slightly different cuts, namely, 10<$E_T^{\text{miss}}$<150 GeV and 10<$E_T^{\gamma}$<150 GeV.
In Fig. \ref{MET} we present the $E_T^{\text{miss}}$ distribution to both signal and background processes, while in figure \ref{ET} we show the photon transverse energy. 
We observe that $E_T^{\gamma}$ and $E_T^{\text{miss}}$ grow as $m_{\chi}$ increase. For a better illustration, Fig. \ref{DMmasses-MET-ET} shows the normalized $E_T^{\gamma}$ and $E_T^{\text{miss}}$ distributions for $m_{\chi}=$10, 100 and 500 GeV.
\subsection{Signal significance}
\begin{figure}[htb!]
\centering
\includegraphics[scale=0.4,angle=270]{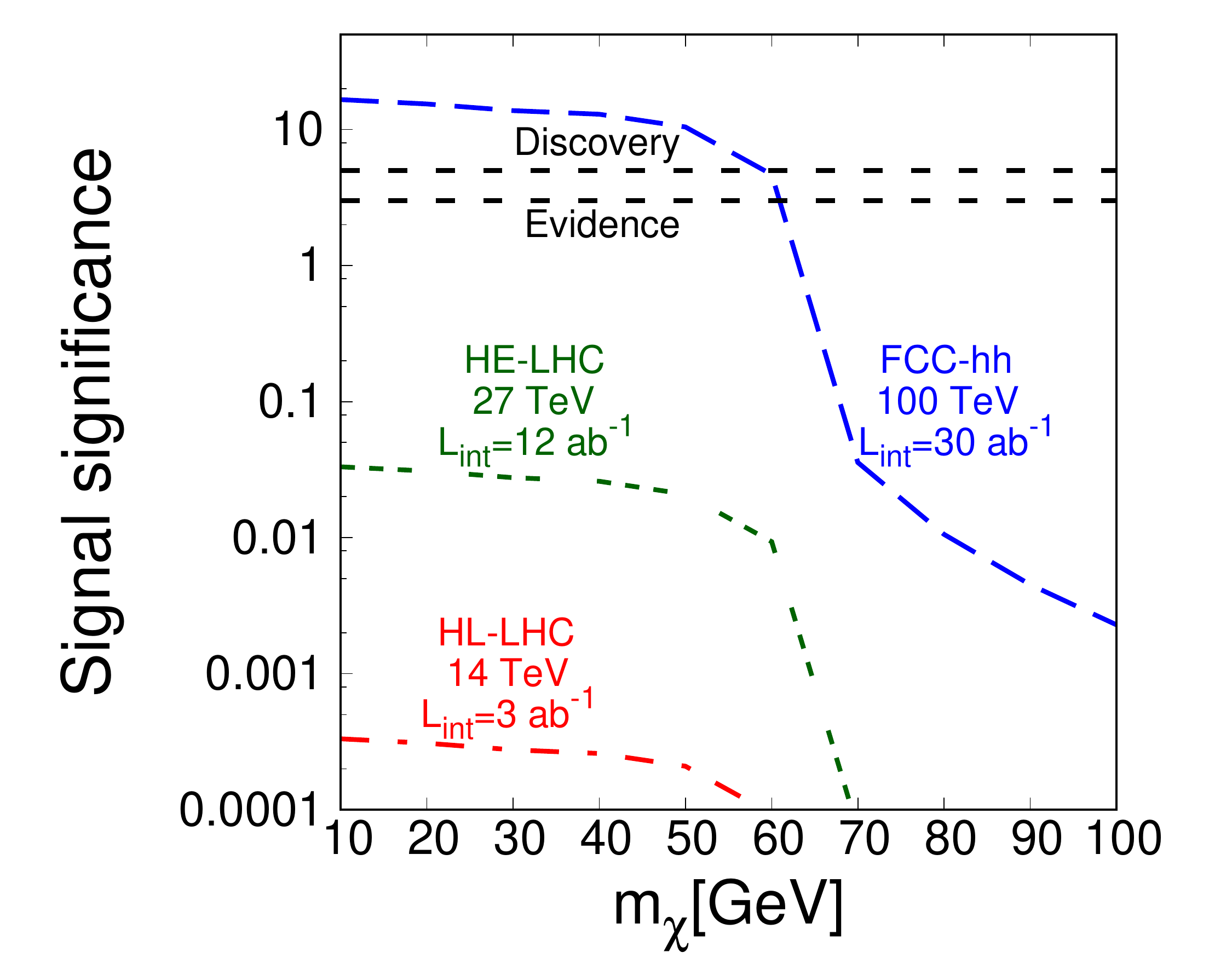}
\caption{Signal significance as a function of the DM candidate mass.
\label{Significance}}
\end{figure}

We compute the signal significance defined as $\textbf{S}=\frac{N_S}{\sqrt{N_S+N_B}}$, where $N_S$ ($N_B$) are the number of signal (background) events after the kinematic cuts were applied. For colliders considered (HL-LHC, HE-LHC and FCC-hh), we find that through $pp\to\chi\chi\gamma$ production only at the FCC-hh will be possible to claim detection of the DM candidate in the range of masses 10-60 GeV once a center-of-mass energy of 100 TeV and an integrated luminosity about 22000 fb$^{-1}$ are reached. This is illustrated in Fig. \ref{Significance} which shows the signal significance as a function of the DM candidate mass.

%
%

%

%
%
\section{Conclusions}
\label{sec7}
In this work, we study an extension of the SM with $U(1)_X$ gauge symmetry that includes two doublets and one complex singlet scalar field in order to introduce a WIMP as DM candidate. The proposed candidate as DM in this extension arises from one inert doublet, whose VEV is equal to zero. In order to ensure the stability of the DM candidate, we consider two scenarios to control the scalar couplings: a discrete $Z_2$ symmetry and the $U(1)_X$  symmetry. 

In the constrained IDMS \cite{Bonilla:2014xba}, the parameters associated with the singlet cubic terms in the scalar potential are responsible for the source of CP violation, since three neutral scalars are mixed to generate the physical states. In the IDMS with local gauge $U(1)_X$ symmetry this type of mixture, which produces an undefined state of CP for neutral scalars, can also occur when $\lambda_6 \neq 0$ and $\lambda_{12x} \neq 0$. However, this analysis is out of the objective at the moment. Then, the study of explicit CP violation can be considered with this model.

We explore the allowed regions for free model parameters of the IDMS taking into account the most up-to-date experimental collider and astrophysical results. We find that the signal strength $\mathcal{R}_{WW^*}$ is the most stringent, allowing an interval for the neutral scalar mixing angle such that $0.99\lesssim \cos{\alpha_{1}}\lesssim 1$. The analysis of the $\sigma(pp\to Z^{\prime})$ production cross-section times $\mathcal{B}(Z^{\prime}\to \ell^-\ell^+)$, with $\ell=e,\,\mu$, excludes regions for $m_{Z^{\prime}}\lesssim3$ TeV with $g_x=0.4$.

Regions for the masses of DM candidate in the order of light ($\mathcal{O}(10)$ GeV), intermediate ($\mathcal{O}(100)$ GeV), and heavy ($\mathcal{O}(2)$ TeV) are in agreement with the upper limit on $\sigma^{SI}(\chi N\to \chi N)$ and relic density reported by XENON1T and PLANCK collaborations, depending mainly on $\lambda_{345}$ and $\lambda_{2x}$. We find that the allowed interval for the DM candidate mass is highly sensitive to $\lambda_{345}$ and $\lambda_{2x}$. For instance, for the values of $\lambda_{2x}=0$ and $\lambda_{345}=0.03$, the allowed values for DM mass are obtained such that $1.1$ TeV$\lesssim m_{\chi} \lesssim 1.6 $ TeV meanwhile for $\lambda_{2x}=0.005$ and $\lambda_{345}=0.01$ the result is $ m_{\chi} \lesssim 0.7 $ TeV.  Additionally, IDMS presents an improvement in the mass region due to the portals associated with the $Z^\prime$ gauge boson and a scalar boson $S$, both portals are predicted by the IDMS, which are absent in models as IDM. 

We conclude that the IDMS is a viable model for the study of DM which provides an improvement in the allowed regions of the DM candidate mass. The IDMS has a rich phenomenology through processes involving $Z^{\prime}$, $S$ and $H^{\pm}$ bosons that could be tested at hadron colliders. Also, the IDMS predict DM particle masses in the interval 10-60 GeV that could be detectable at the FCC-hh through the $pp\to\chi\chi\gamma$ process. However, other processes could also be analyzed to complement the search for DM particles. On the other hand, we also find restrictions for  $\lambda_{4,5}$ parameters of the model that prohibit the decay $\chi\to W^{\pm}H^{\mp}$.

%

 %
\begin{acknowledgements}
M. A. Arroyo-Ure\~na especially thanks to \emph{PROGRAMA DE BECAS POSDOCTORALES DGAPA-UNAM} for postdoctoral funding and thankfully acknowledge computer resources, technical advise and support provided by Laboratorio Nacional de Superc\'omputo del Sureste de M\'exico.
This work was supported by projects \emph{Pro\-gra\-ma de A\-po\-yo a Proyectos de Investigaci\'on e Innovaci\'on
Tecnol\'ogica} (PAPIIT) with registration codes IA107118 and IN115319 in \emph{Direcci\'on General de Asuntos de Personal
Acad\'emico de Universidad Nacional Aut\'onoma de M\'exico} (DGAPA-UNAM), and \emph{Programa Interno de Apoyo para
Proyectos de Investigación} (PIAPI) with registration code PIAPIVC07 in FES-Cuautitl\'an UNAM and \emph{Sistema Nacional de Investigadores} (SNI) of the \emph{Consejo Nacional de Ciencia y Tecnolog\'ia} (CO\-NA\-CYT) in M\'exico.
\end{acknowledgements}

\appendix
\section{Decay widths of scalar $S$ and pseudoscalar $A$ bosons}
\label{App-A}
\subsection{Scalar boson decays}

The most relevant decays of both $CP$-even and $CP$-odd scalar bosons have been long studied in the literature. We will present the decay width formulas for the sake of completeness. The tree-level two-body widths are given as follows:

\begin{eqnarray}
\Gamma(S\to \bar{f}_i f_j)&=&\frac{g_{Sf_if_j}^2N_c m_S}{128\pi}\left(4-(\sqrt{\tau_{f_i}}+\sqrt{\tau_{f_j}})^2\right)^{\frac{3}{2}}\\&\times& \nonumber
(4-(\sqrt{\tau_{f_i}}-\sqrt{\tau_{f_j}})^2)^{1/2},
\end{eqnarray}
with $\tau_i=4m_i^2/m_S^2$ and $N_c$ is the color number. From here we easily obtain  the flavor conserving decay width. 
The $CP$-even scalar boson  decays into pairs of real electroweak gauge bosons can also be kinematically allowed. The corresponding decay width is
\begin{equation}
\Gamma(S\to VV)=\frac{g_{SVV}^2 m_H^3}{64n_V\pi m_V^4}\sqrt{1-\tau_V}\left(1-\tau_V+\frac{3}{4}\tau_V^2\right),
\end{equation}
with $n_V=1\; (2)$ for $V=W\;(Z)$ and  $g_{S\bar{f}_if_j}$-$g_{SVV}$ given in the table \ref{FR-IDMS}.

Additionally to the tree level decays, other relevant channels arise at one-loop, such as $S\to\gamma\gamma$ and $S\to gg$, whose decay widths are given by:
\begin{equation}
\label{Htogammagamma}
\Gamma(S\to \gamma\gamma)=\frac{\alpha^2m_S^3}{1024\pi^3m_W^2}\left|\sum_s A_s^{S\gamma\gamma}\left(\tau_s\right)\right|^2,
\end{equation}
with the subscript $s$ standing for the spin of the charged particle circulating into the loop. The $A_s^{S\gamma\gamma}$ function is given by
\begin{equation}
A_s^{S\gamma\gamma}(\tau_s)=\left\{
\begin{array}{cr}
\sum_f \frac{2m_W g_{Sff} N_c Q_f^2}{m_f}\left[-2\tau_s\left(1+(1-\tau_s)f\left(\tau_s\right)\right)\right]&s=\frac{1}{2},\\ \\
 \frac{ g_{SWW}}{m_W}\left[2+3\tau_W+3\tau_W(2-\tau_W)f\left(\tau_W\right)\right]&s=1,\\ \\
 \frac{m_W g_{SH^+H^-}}{m_{H^-}^2}\left[\tau_{H^{\pm}}\left(1-\tau_{H^{\pm}}f\left(\tau_{H^{\pm}}\right)\right)\right]&s=0,
 \end{array}\right.
\end{equation}
where
\begin{equation}
f(x)=\left\{
\begin{array}{cr}
\left[\arcsin\left(\frac{1}{\sqrt{x}}\right)\right]^2&x\ge1,\\
-\frac{1}{4}\left[\log\left(\frac{1+\sqrt{1-x}}{1-\sqrt{1-x}}\right)-i\pi\right]^2&x<1.
\end{array}
\right.
\end{equation}
The two-gluon decay can only receive contributions from quarks and its decay width can be obtained from \eqref{Htogammagamma} by only summing over quarks and making the replacements $\alpha^2\to 2\alpha^2_S$, $N_c Q_f^2\to 1$.
%
%

%
%
\end{document}